\documentclass[11pt,english]{article}
\pdfoutput=1
\usepackage{jheppub}
\usepackage{amsmath}
\usepackage{amssymb}
\usepackage{amsfonts}

\usepackage{graphicx}

\usepackage{braket}
\usepackage{wrapfig}
\usepackage{subcaption}

\newcommand{\beq}{\begin{equation}}
\newcommand{\eeq}{\end{equation}}

\newcommand{\nn}{{\nonumber}}
\newcommand{\m}{\mu}

\newcommand{\calO}{{\mathcal O}}
\newcommand{\hatomega}{\omega_{-1}}
\newcommand{\hatomegaT}{\omega_{-1}}
\newcommand{\omegac}{\omega_{c_0}}
\newcommand{\omegacT}{\omega_{c_0}}
\newcommand{\D}{\bar{D}}

\begin{document}
\title{Black hole one-loop determinants in the large dimension limit}
\author{Cynthia Keeler}
\author{and Alankrita Priya}
\affiliation{Physics Department,\\
 Arizona State University, Tempe, AZ, 85287, USA}
\emailAdd{keelerc@asu.edu} \emailAdd{apriya@asu.edu}

\abstract{We calculate the contributions to the one-loop determinant for transverse traceless gravitons in an $n+3$-dimensional Schwarzschild black hole background in the large dimension limit, due to the $SO(n+2)$-type tensor and vector fluctuations, using the quasinormal mode method. Accordingly we find the quasinormal modes for these fluctuations as a function of a fiducial mass parameter $\Delta$.  We show that the behavior of the one-loop determinant at large $\Delta$ accords with a heat kernel curvature expansion in one lower dimension, lending further evidence towards a membrane picture for black holes in the large dimension limit.}

\maketitle


\section{Introduction}   
Quantum effects in nontrivial gravitational backgrounds are of great interest to the theoretical physics community. Even the leading one-loop effects can contain important physical results, such as quantum corrections to the entropy of black holes, which a series of papers \cite{Mandal:2010cj,Sen:2008vm, Banerjee:2011jp,Liu:2017vbl,Murthy:2015yfa,Chowdhury:2014lza,Gupta:2014hxa,Gupta:2013sva,Banerjee:2010qc,Sen:2008yk} found via calculations of one-loop determinants. 

Since the computation of one-loop determinants, and thus one-loop partition functions, is technically difficult in generic curved spacetimes, several methods have been developed to handle the computations.  There are three primary strategies: heat kernel methods, group theoretic approaches, and the quasinormal mode method.  Heat kernel methods may be exact, as in the eigenfunction expansion (e.g. \cite{Camporesi:1991nw,CAMPORESI199457,Camporesi:1992wn,Camporesi:1994ga}) or the method of images (e.g. \cite{Giombi:2008vd}); alternatively, the heat kernel curvature approximation is only appropriate for fluctuations of massive fields (see \cite{
Vassilevich:2003xt} for a review of the heat kernel approach). 
For spacetimes with a simple symmetry structure, physicists have applied two group theoretic approaches.  First, the authors of \cite{David:2009xg,Gopakumar:2011qs} use characters of representations to build the explicit expression for the one loop determinant of a fields with arbitrary spin in odd-dimensional AdS spaces.   In the same spirit, the authors of \cite{Larsen:2014bqa,Larsen:2015aia, 1505.01156,Keeler:2014bra} use the group theoretical structure of black hole solutions to predict their one loop determinants.

Around a decade ago, in two papers \cite{Denef:2009kn,Denef:2009yy}, Denef, Hartnoll and Sachdev developed a method  for the computation of one-loop determinant not directly based on the heat kernel approach. The key insight of \cite{Denef:2009kn} is to build the one-loop determinant as a function of mass parameter $\Delta$ using quasinormal frequencies of the field fluctuations in the bulk. These frequencies turn out to be poles (zeros) of the one-loop determinant for bosonic (fermionic) field fluctuations. A series of papers have employed this approach for one-loop calculations \cite{Castro:2017mfj, Keeler:2016wko, Keeler:2014hba, Zhang:2012kya, Arnold:2016dbb}. We will use this method to compute the one-loop determinant for gravitational perturbations around the Schwarzschild black hole in the large dimension limit.

 In a finite number of dimensions, the Einstein equation lacks a small parameter; in the limit where the number of dimensions $D$ is taken large, then $1/D$ can provide a perturbative parameter \cite{Emparan:2013moa}.  For the Schwarzschild black hole in a small number of dimensions, there is no separation of scales between the horizon dynamics and the asymptotic behavior.  Conversely, in the large $D$ limit, the gravitational field of a black hole becomes strongly localized near the horizon, effectively decoupling the black hole dynamics from the asymptotic structure. As suggested in  \cite{Bhattacharyya:2015dva,Emparan:2015hwa}, we can think of the dynamics of a such a large dimension black hole as equivalent to those of a membrane propagating in flat space.  Since these developmental works, there has been significant interest in black hole spacetimes in the large dimension limit \cite{Emparan:2013moa,Emparan:2014cia,Emparan:2014aba,Emparan:2015hwa,Dandekar:2016fvw,Tanabe:2016opw,Saha:2018elg,Chen:2017wpf,Bhattacharyya:2016nhn,Rozali:2016yhw,Emparan:2015rva,Bhattacharyya:2015fdk,Bhattacharyya:2015dva,Suzuki:2015iha,Herzog:2017qwp,Mandlik:2018wnw,Emparan:2014jca,Rozali:2018yrv,Andrade:2018nsz,Andrade:2018rcx,Bhattacharyya:2017hpj}.

Of particular relevance for our work,  \cite{Emparan:2014aba} computes the spectrum of massless quasinormal modes in a $1/D$ expansion. Using the gauge-invariant formalism of \cite{Kodama:2003jz, Kodama:2000fa}, the authors of \cite{Emparan:2014aba} find two distinct sets of quasinormal modes.  First, they consider modes whose frequencies are of order $D$, finding that the dynamics of these modes depends mostly on the flat asymptotics of the spacetime.  The second sequence of modes has frequencies of order $D^0$ and lives entirely in the near-horizon region.  Hence, the dynamics of these decoupled modes reflects the fluctuations of the near-horizon membrane region. As we will show, it is this second decoupled sector of modes which are of greatest importance in the one-loop determinant.

We compute the one loop determinant for gravitational fluctuations in the membrane region of the large dimension Schwarzschild black hole, via the quasinormal mode method. Like \cite{Emparan:2014aba}, we adopt the gauge-invariant formalism of Kodama+Ishibashi \cite{Kodama:2003jz, Kodama:2000fa} to compute the quasinormal frequencies, now as a function of a mass parameter $\Delta$. We consider only transverse traceless fluctuating modes, as they have been successful previously for 3 dimensional gravitons in AdS$_3$ \cite{Castro:2017mfj}, combined with appropriate gauge fixing, in obtaining the one loop determinant for graviton fluctuations in Einstein gravity.  We emphasize that these transverse traceless gravitons, often referred to as `tensor' fluctuations due to their behavior under the local Lorentz symmetry, further break down into tensor, vector and scalar components with respect to the spherical symmetry of the spacetime.
Accordingly we use the terminology `scalar', `vector', and `tensor' to refer to this spherical symmetry, and not to the full local Lorentz symmetry (as is otherwise common, e.g. in \cite{VanNieuwenhuizen:1973fi}).
 At $\Delta$ of order $D^0$, we find that only the decoupled, membrane-region quasinormal modes provide mass-plane poles. There are no contributing modes in the tensor sector,  while in the vector sector the poles are simply defined by the quantum numbers, at least at leading order in $D$. Using these poles, we build the one-loop determinant for vector modes and express it in terms of Hurwitz zeta function. We find that the leading behavior of the one-loop determinant is proportional to $\Delta^{D-1}$. This dimensionality reduction in the exponent corresponds with the membrane paradigm picture of black holes in the large dimension limit. We also compare our results with heat kernel curvature calculations.

The paper is outlined as follows: In Sec.~\ref{sec:review} we review the large dimension limit of the Schwarzschild black hole and the quasinormal mode method. In Sec.~\ref{QNM3} we compute the quasinormal modes for the graviton field with an added fiducial mass. In Sec.~\ref{4} we write the expression for the one-loop determinant in terms of elementary zeta functions, and compare this expression with the heat kernel curvature calculation. In Sec.~\ref{5} we conclude with some comments and open questions.

\section{Review}\label{sec:review}
In this section we review the large dimension limit (Section \ref{sec:reviewLarge}) and the quasinormal mode method (Section \ref{sec:reviewQNM}) and set our notation.  The reader familiar with both topics may wish to skip to Section \ref{QNM3}.

\subsection{The large dimension limit}\label{sec:reviewLarge}

We review the large dimension limit of a Schwarzschild black hole, as exhibited in \cite{Emparan:2014aba} and \cite{Bhattacharyya:2015dva}. Since our goal will be to calculate quasinormal mode frequencies for massive modes, we will largely follow the method and notation of \cite{Emparan:2014aba}, although we will also draw upon the conceptual approach in \cite{Bhattacharyya:2015dva}. 
The metric for a Schwarzschild black hole of radius $r_0$ in $D$ total dimensions is
\begin{equation}\label{metriceqn}
ds^2=-f(r)dt^2+\frac{dr^2}{f(r)}+r^2d\Omega_{n+1}, \hspace{3mm} f(r)=1-\left(\frac{r_0}{r}\right)^{n},
\end{equation}
where $n=D-3$. As we can see from the metric, in the large $D$ (or $n$) limit there are two important regions outside the horizon, with markedly different behavior.  First, for any $r>r_0$, with $r_0$ held fixed as $D\rightarrow\infty$, $f(r)\rightarrow1$, so the metric reduces to flat space. We refer to this region as the `far' region.

On the other hand, if we examine a very near-horizon region by setting $r=r_0\left(1+\frac{\lambda}{D-3}\right)$ and instead keeping $\lambda$ fixed as $D\rightarrow\infty$, then $f(r)\rightarrow1-e^{-\lambda}$. Thus the nontrivial gravitational field, where $f(r)$ is substantially different from one, is strongly localized in a thin near-horizon region of thickness $\sim r_0/D$. This near-horizon `membrane' region encodes the most important black hole physics \cite{Bhattacharyya:2015dva,Emparan:2015hwa}.

\begin{figure}[t]
\begin{center}
\includegraphics[width=0.8\linewidth]{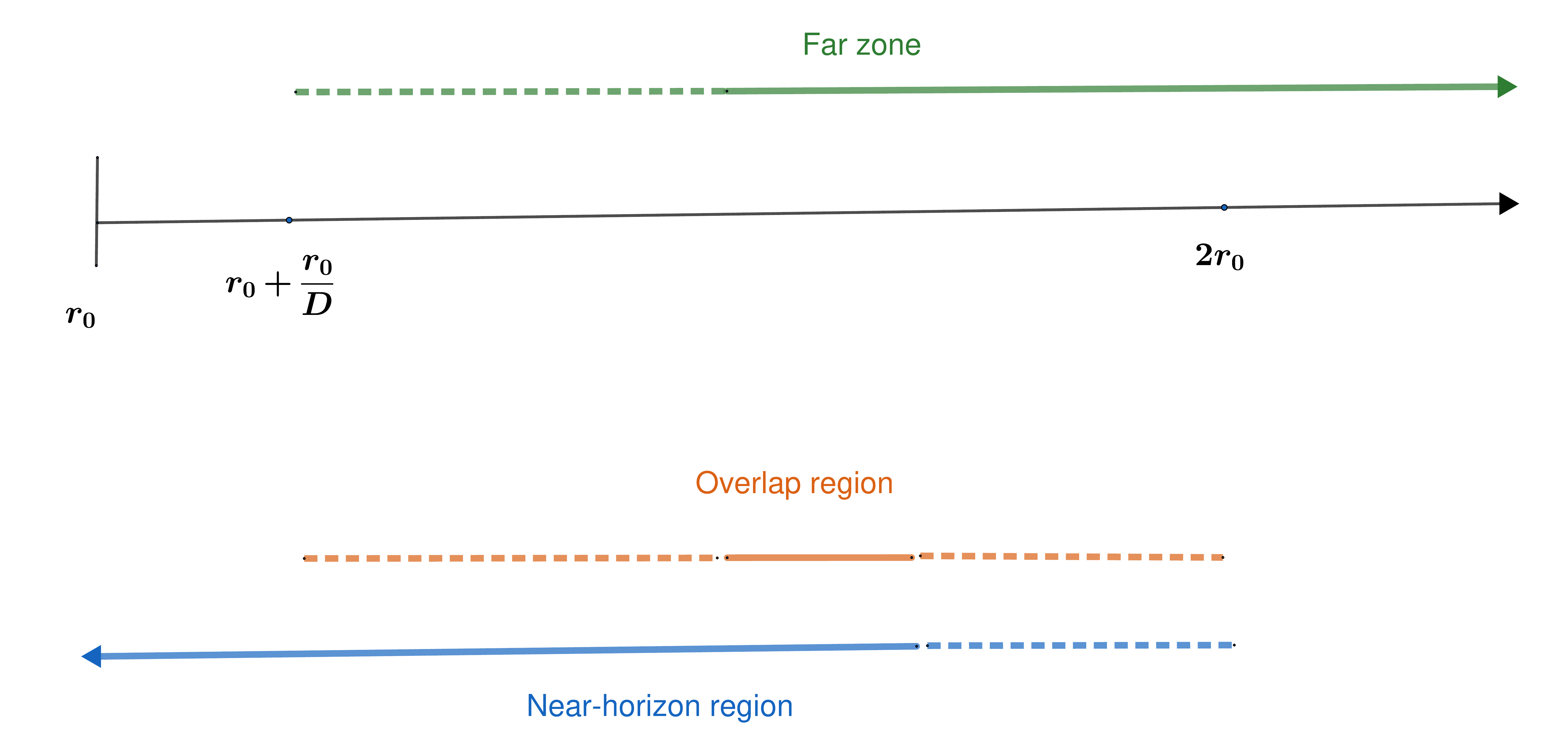}
\end{center}
\caption{A schematic diagram representing the range of the near-horizon, far zone and overlap regions.\label{NearVSFar}}
\end{figure}

In the near-horizon zone, we will use the radial coordinate $\rho=\left(r/r_0\right)^n$ where the metric becomes
\begin{equation}\label{NHmetric}
ds^2=-\left(1-\frac{1}{\rho}\right) dt^2+\frac{r_0^2}{n^2}\frac{d\rho^2}{\rho(\rho-1)}+r_0^2d\Omega_{n+1}.
\end{equation}\
The near and far regions actually overlap, as depicted in Figure \ref{NearVSFar}:
\begin{equation}
\mathrm{near\hspace{1mm} region}: r-r_0<<r_0\\
; \hspace{2mm}\mathrm{far\hspace{1mm} region}: r-r_0>>r_0/D
\end{equation}
Or, equivalently in terms of $\rho$:
\begin{equation}\label{rhoregion}
\mathrm{near\hspace{1mm} region}: \ln \rho<< D\\
; \hspace{2mm}\mathrm{far\hspace{1mm} region}: \ln \rho>>1
\end{equation}
The overlap region is thus
\begin{equation}
r_0/D<<r-r_0<<r_0, \hspace{2mm} \mathrm{or}, \hspace{1mm} 1<<\ln \rho<<D.
\end{equation}

\cite{Emparan:2014aba} studied the linearized gravitational perturbations around the Schwarzschild black hole in the large dimension limit, and computed the quasinormal spectrum of its oscillations in analytic form in the $1/D$ expansion. They found two distinct sets of modes:
\begin{itemize}
\item Non-decoupled modes (heavy), with frequencies of order $D/r_0$, lying between the near-horizon and asymptotic region. Most quasinormal modes fall in this class but they carry very little information about the black hole and so are of least physical interest.
\item Decoupled modes (light), with frequencies of order $1/r_0$, are decoupled from the asymptotic region and are localized entirely inside the membrane region. There are only three such modes (two scalars and one vector)\footnote{Note that the scalar/vector here refers to the decomposition with respect to the spherical symmetry of the spacetime as used in the gauge invariant formalism of Kodama+Ishibashi \cite{Kodama:2003jz, Kodama:2000fa}.}.
\end{itemize}
The decoupled modes capture the interesting physics specific to each black hole, such as stability properties.  Accordingly, we will choose a limit that focuses on these decoupled modes. In \cite{Emparan:2014aba}, the authors recover both sets of modes by studying the linearized gravitational perturbations $h_{\mu\nu}=e^{-i\omega t}h_{\mu\nu}(r,\Omega)$ of the metric \eqref{metriceqn} under the vacuum GR equations. The modes can additionally be separated according to their angular dependence; that is, via their transformation properties under the $SO(n+2)$ symmetry of the $S^{n+1}$ sphere.  There are thus three types of linearized graviton modes: scalar-type ($S$), vector-type ($V$) and tensor-type ($T$).%
\footnote{Note this mode separation is not equivalent to separating under the local Lorentz symmetry, such as used in \cite{Yasuda:1983hk}. We will return to this issue in the discussion in Section \ref{5}.} 
Following \cite{Kodama:2003jz}, we can study each of these modes in terms of a single gauge invariant master variable $\psi(r)$ that satisfies master equations of the form
\begin{equation}\label{mastereqn}
\frac{d}{d r}\left(f \frac{d\psi}{dr}\right)-\frac{V_s \psi}{f}+\frac{\omega^2}{f}\psi=0
\end{equation}
where $s$ stands for either Tensor(T), Vector(V) or Scalar(S). The effective tensor potential is given by
\begin{equation}\label{2.7}
V_T=\frac{n^2 f}{4 r^2}\left[\left(1+\frac{2l}{n}\right)^2-\frac{1}{n^2}+\left(1+\frac{1}{n}\right)^2\left(\frac{r_0}{r}\right)^n\right],
\end{equation}
where $l$ is the conserved angular momentum from the $SO(n+2)$ symmetry.
For vectors the effective potential is
\begin{equation}\label{2.8}
V_V=\frac{n^2 f}{4 r^2}\left[\left(1+\frac{2l}{n}\right)^2-\frac{1}{n^2}-3\left(1+\frac{1}{n}\right)^2\left(\frac{r_0}{r}\right)^n\right].
\end{equation}
The scalar potential can be found in the Appendix in equation \ref{scalarpotential}.

When finding the quasinormal modes, as done in \cite{Emparan:2014aba}, the key idea is to first solve the linearized field equation \eqref{mastereqn} perturbatively in the $1/n$ expansion in the near region, with ingoing boundary conditions imposed at the horizon. Next the equation should be solved in the far region, with outgoing or normalizable boundary conditions at infinity.  Last, the near and far region solutions should be matched in the overlap region.

\begin{figure}[h]
\begin{center}
\includegraphics[width=0.5 \linewidth]{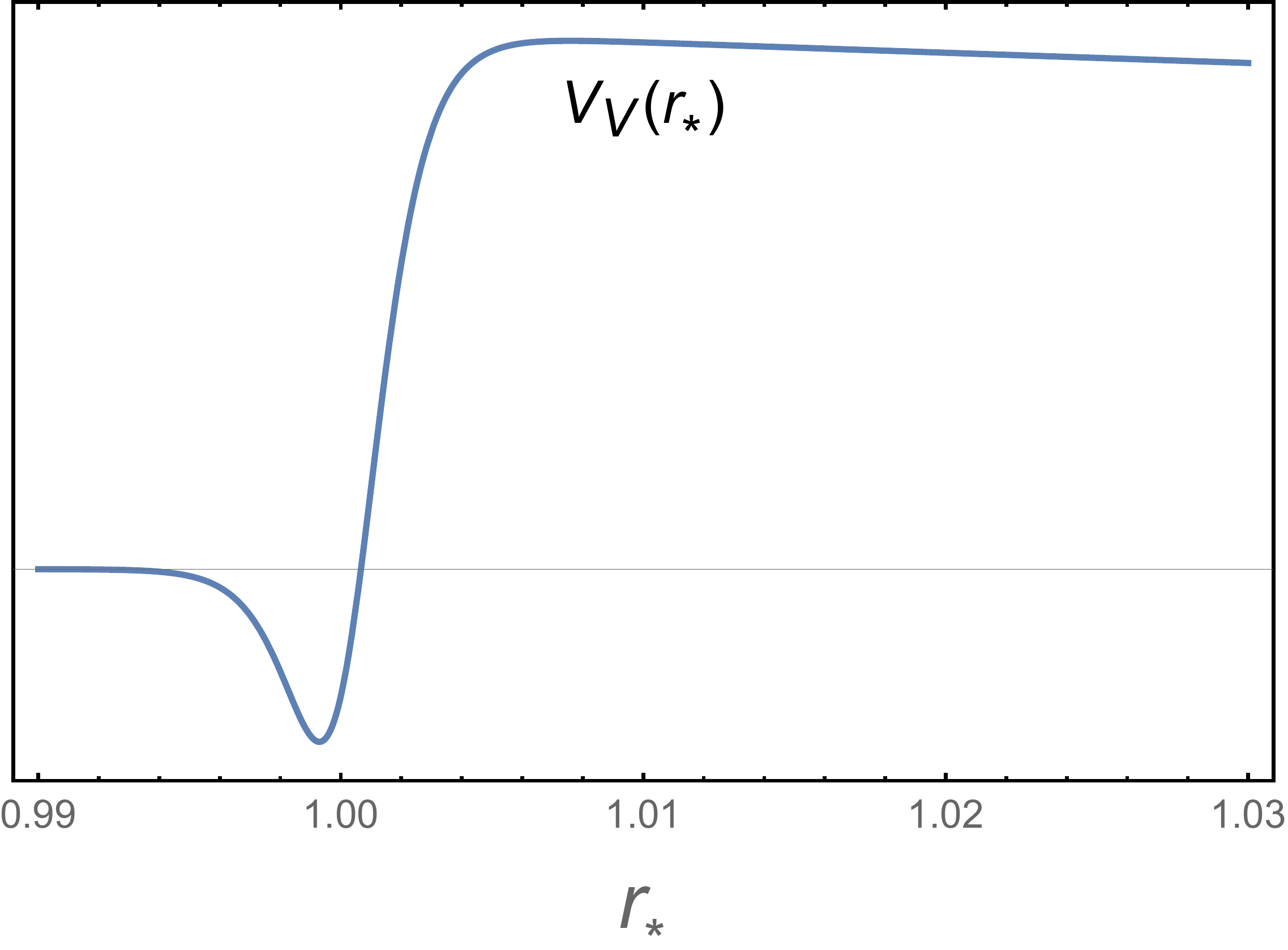}
\end{center}
\caption{\label{figure1}A schematic diagram representing the vector potential $V_V(r_*)$ in the near-horizon region for $n=D-3=1000$ and angular momentum $l=5$, where $r_*=\int dr/f $. The decoupled (light) modes of frequency $\omega \sim  \mathcal{O}(1)$ in $n$ live in the dip of the potential.}\label{VectorPot}
\end{figure}

We can understand the distinction between the near-horizon, decoupled modes and the far, coupled modes by examining the form of these potentials.  In Figure \ref{VectorPot}, we present an example of one such potential, for a fixed but large value of the dimension, $D-3=n=1000$, and a representative value of $l=5$ for the angular momentum.
The lightest modes of $\mathcal{O}(n^0)$ frequency will live in the near-horizon dip of the potential.  These modes are the decoupled modes.  At large frequencies, another set of modes lives exclusively in the far region, at larger radius than the maximum of the potential.

Again, our primary interest will be in the physics of the black hole itself, so we will concentrate on the light, decoupled, near-horizon modes.  As we will see in the next section, we need to extend the calculation of \cite{Emparan:2014aba} to include a formal mass for the graviton.

\subsection{Quasinormal mode method}\label{sec:reviewQNM}
In this section we review the quasinormal mode method for computing one-loop determinants as developed in \cite{Denef:2009kn}. For simplicity, we present the method here for a complex scalar, although we will use it for gravitons in our work below.

For bosonic degrees of freedom, the one-loop determinant for a complex scalar field $\psi$ of mass $m$ is
\begin{equation}\label{pathint}
Z^{(1)}(m^2)=\int D\psi e^{-\int d^2x \psi^{*}(-\nabla^2+m^2)\psi} \propto \frac{1}{\mathrm{det}(-\nabla^2+m^2)},
\end{equation}
where $\nabla^2$ is the kinetic operator on the given spacetime background%
\footnote{As in \cite{Denef:2009kn} this method formally applies to Euclidean spacetimes, so we are really calculating (a portion of) the one-loop determinant for the Euclidean thermal spacetime that is the Wick-rotation of \eqref{metriceqn}, even though we will do so using the Lorentzian information of quasinormal modes.}%
. 
If the background is non-compact, then we examine the boundary conditions we want to impose on fields at infinity, and choose a mass parameter $\Delta$, a function of the mass $m^2$, accordingly.  For the case of AdS backgrounds, this parameter simply becomes the conformal dimension, hence the notation $\Delta(m^2)$; more generally it is chosen so the boundary conditions depend meromorphically on $\Delta$.
 
Next, we assume that $Z^{(1)}(\Delta(m^2))$, analytically continued to the $\Delta$ complex plane, is a meromorphic function%
\footnote{This assumption is unproven, but has been successful thus far. Meromorphicity in the parameter $s$, of zeta functions of the form $\zeta_A(s)=Tr(A)^{-s}$, for self-adjoint operators $A$, has been studied extensively \cite{Seeley:1967ea}, due to its use in the zeta function regularization method \cite{Dowker:1975tf,Hawking:1976ja}; however, these works do not specifically study meromorphicity in a mass parameter (which would change the operator $A$).  Additionally, we emphasize that the QNM method is \emph{not} equivalent to zeta function regularization, although it is closely related; for more on the multiplicative anomaly that the QNM method resolves in a fundamentally different manner, see e.g. \cite{Cognola:2014pha} .}
.  This assumption allows us to use the Weierstrass factorization theorem to find the determinant.   As a consequence of this theorem, a meromorphic function is determined (up to a single \emph{entire} function) by the locations and degeneracies of its zeros and poles. As we can see from the right hand side of \eqref{pathint} that the determinant for bosonic degrees of freedom will have no zeros, so we can write
\begin{equation}\label{scalardetform}
Z^{(1)}(\Delta)=e^{\mathrm{Pol}(\Delta)}\prod\limits_{i}(\Delta-\Delta_i)^{-d_i}                         
\end{equation}
where the $\Delta_i$ are the poles (of $Z^{(1)}(\Delta)$) each with degeneracy $d_i$. $\mathrm{Pol}(\Delta)$ is a polynomial in $\Delta$, since $e^{\mathrm{Pol}(\Delta)}$ is entire (that is, it is meromorphic and has no zeros or poles itself). 

Poles occur when $\det=0$, which happens when the mass parameter $\Delta(m^2)$ is set so there is a $\psi$ that solves the Klein-Gordon equation $(-\nabla^2+m^2)\psi=0$ while simultaneously being smooth and regular everywhere in the Euclidean background space. \cite{Denef:2009kn} shows that for static Euclidean thermal spacetimes, the allowed zero modes Wick rotate to (anti)quasinormal modes of the corresponding Lorentzian black holes with frequencies satisfying
\begin{equation}\label{QNMcondition}
\omega(\Delta_i)=2\pi i pT,
\end{equation}
where $p\in\mathbb{Z}$. For $p\geq0$, the Euclidean modes match onto ingoing quasinormal modes, while for $p\leq0$, they instead match onto outgoing quasinormal modes (or anti-quasinormal modes).  Consequently, if we know the (anti)quasinormal mode frequencies as a function of mass for a complex scalar field in a black hole background, then we can immediately write down the one-loop determinant, up to an entire function, as an infinite product.

Once all poles are determined, we take the logarithm of both sides and use zeta functions to perform the infinite sum. 
If we can characterize the large $\Delta$ behavior of the infinite sum, then we can determine $\mathrm{Pol(\Delta)}$ by matching the large $\Delta$ behavior of the right hand side of \eqref{scalardetform} to the local heat kernel curvature expansion in $1/m$ for $Z(\Delta)$ (see e.g. \cite{Vassilevich:2003xt}).

For our calculation, we will introduce a fictitious mass for the graviton (e.g. following \cite{Castro:2017mfj, Keeler:2018lza, Datta:2011za}) in order to facilitate this mathematical trick.

\section{Calculating the quasinormal modes}\label{QNM3}
\label{sec:QNMs}
In this section we extend the quasinormal mode calculations done in \cite{Emparan:2014aba} by adding a fictitious mass to the graviton field%
\footnote{We must warn the readers that we are not doing massive gravity. The reason for adding a mass term is just to use the quasinormal mode method \cite{Denef:2009kn} as discussed in Section \ref{sec:reviewQNM}.}. %
Here we consider only transverse traceless fluctuations, as they have been successful previously for 3 dimensional gravitons in AdS$_3$ \cite{Castro:2017mfj}.
Our goal is to obtain the quasinormal mode spectrum as a function of this fictitious mass; then we will use the relation \eqref{QNMcondition} to find the corresponding poles in the one loop determinant, in Section \ref{sec:QNMresults}.

\subsection{Setting up the equations}\label{3.1}
We want to study the transverse traceless metric perturbation around our background \eqref{metriceqn} using the quasinormal mode method.  Accordingly we add a fiducial mass, resulting in the action

\begin{equation}
S=\int d^{4}x \sqrt{-g}(R-\frac{1}{4}m^2h_{\mu\nu}h^{\mu\nu}),
\end{equation}
where $m^2$ is the added mass. Additionally, $g_{\mu\nu}=g^0_{\mu\nu}+h_{\mu\nu}$, where $g^0_{\mu\nu}$ is the fixed background metric and $h_{\mu\nu}$ is the metric perturbation. We consider these perturbations to be transverse and traceless. Varying the action with respect to $h^{\mu\nu}$ gives the linearized field equation
\begin{equation}\label{3.2}
\delta R_{\mu\nu}-\frac{R^0}{2}h_{\mu\nu}- g^0_{\mu\nu}\delta R+\frac{m^2}{2}h_{\mu\nu}=0,
\end{equation}
where $\delta R_{\mu\nu}$ and $\delta R$ are taken from Appendix B of \cite{Kodama:2000fa}.

Following \cite{Kodama:2003jz}, we study \eqref{3.2} for three different decompositions of the transverse traceless gravitational perturbation $h_{\mu\nu}$, depending on the transformation properties under the $SO(n+2)$ symmetry of the $S^{n+1}$ sphere: scalar-type ($S$), vector-type ($V$) and tensor type ($T$).

In this paper we will focus on calculations for vector and tensor type perturbations; we partially address the scalar perturbations in Appendix \ref{C}. For tensor type perturbations, the equation of motion \eqref{3.2}, in terms of the master variable $\Psi(t,r)=\psi(r)e^{-i\omega t}$ as defined above equation \ref{PsiTensor}, becomes
\begin{equation}\label{3.3}
\frac{d}{d r}\left(f \frac{d\psi}{dr}\right)-\frac{V_T \psi}{f}+\frac{\omega^2}{f}\psi=m^2 \psi.
\end{equation}
For vector type perturbations, under a parameterization choice we define in Appendix \ref{A}, \eqref{3.2} similarly reduces to
\begin{equation}\label{3.4}
\frac{d}{d r}\left(f \frac{d\psi}{dr}\right)-\frac{V_V \psi}{f}+\frac{\omega^2}{f}\psi-m^2\psi=0.
\end{equation}
The $V_s$ are defined in \eqref{2.7} and \eqref{2.8}. For both the tensor and vector modes, we follow the basic idea of \cite{Emparan:2014aba}, solving these field equations perturbatively order by order in the $1/n$ expansion in order to compute the quasinormal modes.

We choose the convenient mass parameter 
\begin{equation}\label{52c}
\Delta=-\frac{n}{2}+\sqrt{m^2+\frac{n^2}{4}},
\end{equation} 
because our boundary conditions on the field will be analytic in $\Delta$, as we discuss below.  We additionally introduce the parameter $\mu=\sqrt{m^2/n^2+1/4}$, which satisfies the relations
\beq
\Delta=n(\mu-1/2),  \qquad 4m^2=4\mu^2n^2-n^2.
\eeq
Following the quasinormal mode method, we wish to study the quasinormal mode frequencies in the complex $\Delta$ plane.
 
\begin{figure}[h]
\begin{minipage}[c]{.45\linewidth}
\begin{subfigure}[b]{\linewidth}
\centering
\includegraphics[width=\linewidth]{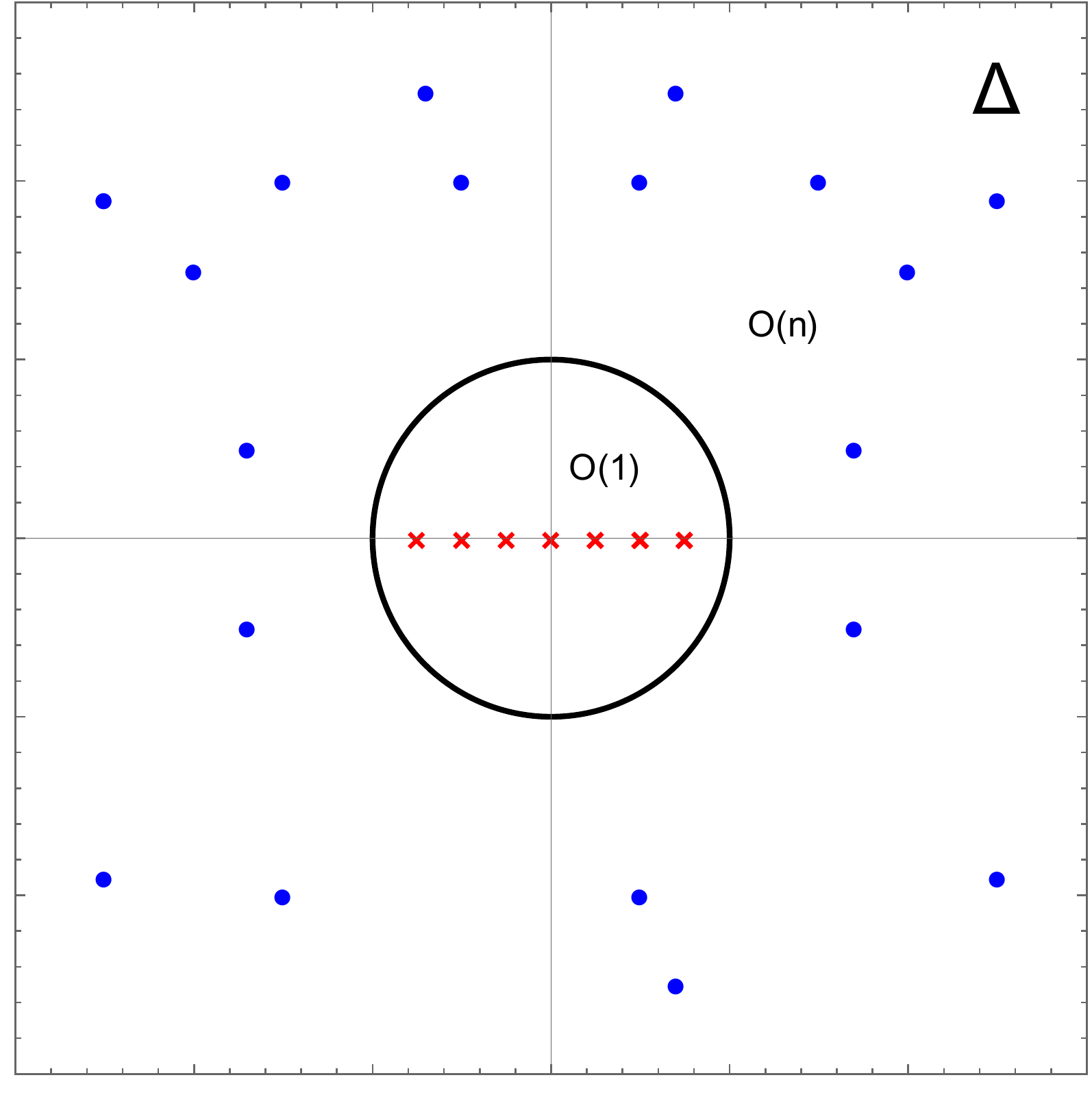}
\caption{A cartoon of poles in the $\Delta$ plane.\label{fig:largeDeltaplane}}
\end{subfigure}
\end{minipage}
\hfill
\begin{minipage}[c]{.45\linewidth}
\begin{subfigure}[b]{\linewidth}
\centering
\includegraphics[width=\linewidth]{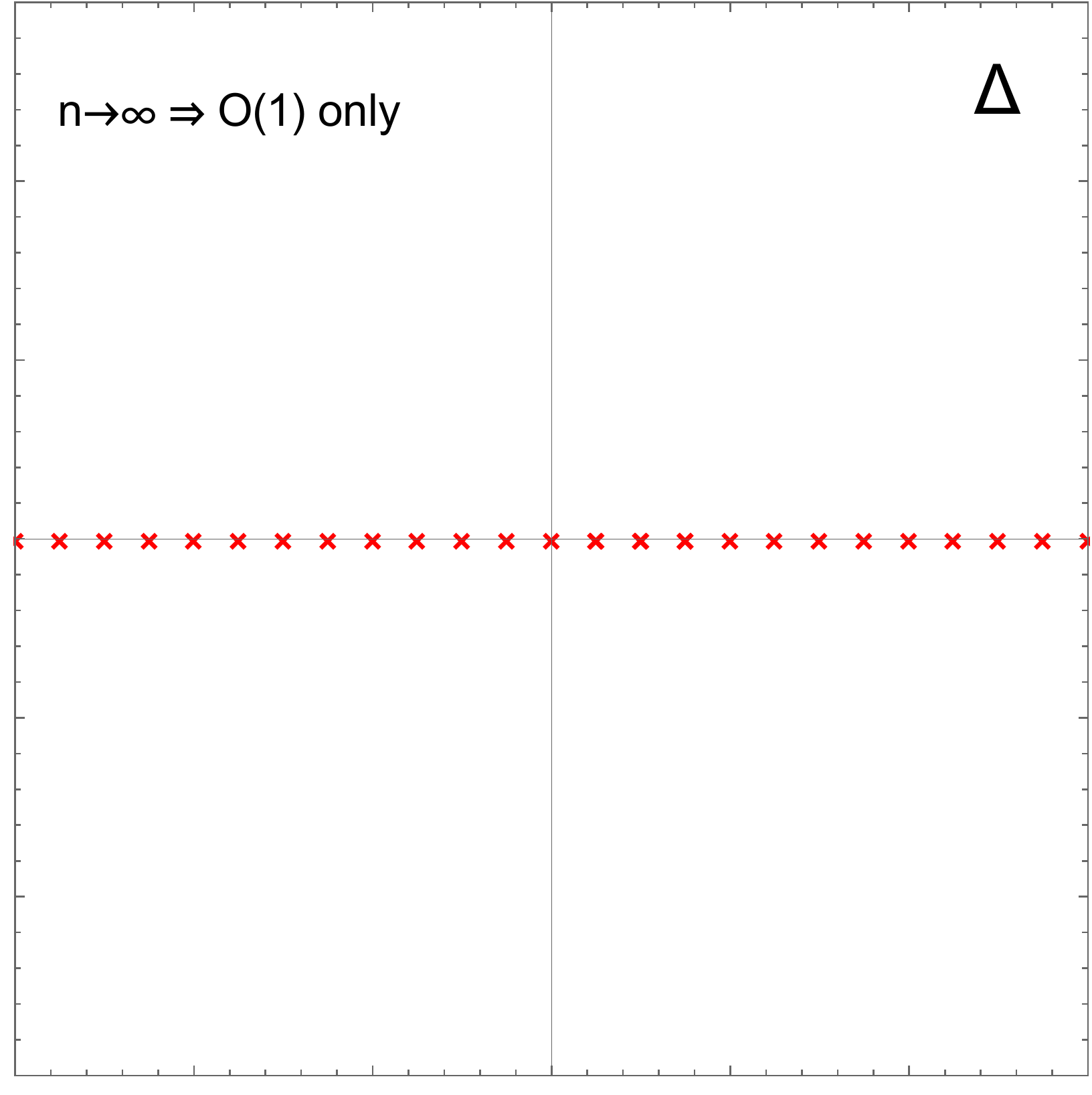}
\caption{A cartoon of the poles in the large $D=n-3$ limit of the $\Delta$ plane.}\label{fig:limitlargeDeltaplane}
\end{subfigure}
\end{minipage}
\end{figure}

Since we are interested in only the near-horizon contribution to the one-loop determinant, we expect that the poles of interest should satisfy $\Delta =\mathcal{O}(n^0)$. Our argument can be understood from figures \ref{fig:largeDeltaplane} and \ref{fig:limitlargeDeltaplane}.%
%
Figure \ref{fig:largeDeltaplane} shows a cartoon of the poles for finite $n$ in the complex $\Delta$ plane. The poles inside the circle are $\mathcal{O}(1)$ and the ones outside are $\mathcal{O}(n)$ and higher. As we increase $n$, the circle will get bigger and $\mathcal{O}(n)$ poles are pushed further out. In the $n\rightarrow \infty$ limit, as shown in \ref{fig:limitlargeDeltaplane}, there will only be $\mathcal{O}(1)$ poles left. Accordingly, in the large $n$ limit we will only be interested in the $\mathcal{O}(1)$ poles; any poles that scale faster in $n$ have been pushed to the edge of the complex plane. Since we only assume our function $Z(\Delta)$ is meromorphic on the complex delta \emph{plane} and not on the full Riemann sphere, we recognize that these poles should not be counted as poles on the $\Delta$ plane in the large $n$ limit.  Rather their effect should be accounted for by the entire function $\exp (\text{Poly})$, which we fix by other means.%
\footnote{This accumulation of poles at $\Delta =\infty$ on the Riemann sphere produces an essential singularity there, but the term $\exp( \text{Poly}(\Delta))$ already provides a way to account for them.}
Although we have not provided a rigorous mathematical proof, our intuition here accords with the physical picture of the large dimension limit, $n\rightarrow\infty$, as taking a near-horizon limit. These $\mathcal{O}(1)$ poles, as discussed in later sections, correspond to modes that live entirely in the near-horizon region and thus capture the relevant physics of interest in the near-horizon limit.

We now wish to study the equations \eqref{3.3} and \eqref{3.4} to solve for quasinormal modes that indicate poles at $\Delta = \mathcal{O}(1)$. Since $\Delta=n(\mu-1/2)$, we can rewrite the requirement for $\Delta$ to be $\mathcal{O}(1)$ as the requirement that $\mu$ satisfy 
\begin{equation}\label{3.5}
\mu=\mu_0+\sum\limits_{k=1}\frac{\mu_k}{n^k}= \frac{1}{2}+\sum\limits_{k=1}\frac{\mu_k}{n^k}.
\end{equation}
In other words, we want to find quasinormal modes whose frequencies satisfy the regularity condition equation \eqref{QNMcondition} when the mass parameter $\mu=1/2 + \mathcal{O}(1/n)$.
As we show below, these modes will primarily live in the near-horizon region, so we now rewrite our equations in terms of the $\rho=(r/r_0)^n$ coordinate from the near-horizon metric \eqref{NHmetric}.  Note that these equations are still exact, in that we have not yet taken either a near-horizon nor a large-dimension limit.

In this near-horizon coordinate $\rho$, and additionally setting $r_0=1$, the tensor and vector equations \eqref{3.3} and \eqref{3.4} become
\begin{equation}\label{3.7}
\left(\mathcal{L}+U_s+\left(1-\frac{1}{\rho}\right)\left(\frac{\Delta}{n}+1\right)\frac{\Delta}{n}\right)\psi=0,
\end{equation}
where $s$ is set to either $T$ or $V$ as appropriate.
Here we have introduced
\begin{equation}
\mathcal{L}=-\frac{(\rho-1)}{\rho^{1/n}}\frac{d}{d\rho}\left(\frac{(\rho-1)}{\rho^{1/n}}\frac{d\psi}{d\rho}\right), \qquad 
U_T=\frac{V_T-\omega^2}{n^2},\qquad \text{and} \qquad U_V=\frac{V_V-\omega^2}{n^2}.
\eeq

In order to find the quasinormal modes, we also need to know their boundary conditions; 
$\psi$ should satisfy the ingoing horizon boundary condition near the horizon, $\rho=1$.  It is easiest to express the ingoing condition in the radial coordinate $r_{*}=\frac{1}{n}\ln(\rho-1)$.  In this coordinate, the horizon is at $r_{*}=-\infty$.  For an ingoing wave, we should expect
\beq
\Psi(r_*,t)=\psi(r_*)e^{-i\omega t} \underset{r_*\rightarrow\infty}{\propto} e^{-i\omega t -i\omega r_*}.
\eeq
The $\propto$ here indicates that the function can have a regular component as it approaches the horizon; since we do not care about the overall normalization of $\psi$ we are free to insist this component approach $1$ at the horizon. Rewriting the horizon boundary condition again in our near-horizon coordinate $\rho$, we have
\begin{equation}\label{quasiBC}
\psi(\rho)\underset{\rho\rightarrow 1}{\propto}(\rho-1)^{-i \omega/n}.
\end{equation}
Anti-quasinormal modes have the same condition except for a change of sign in the exponent:
\begin{equation}\label{antiquasiBC}
\psi(\rho)\underset{\rho\rightarrow 1}{\propto}(\rho-1)^{i \omega/n}.
\end{equation}

Quasinormal modes must also satisfy the no-incoming-flux condition at large radius.  Here we closely follow the analysis done in \cite{Emparan:2014aba}.  In the far zone \eqref{rhoregion}, any term of the form $1/\rho$ is exponentially small in $n$.  Accordingly, we can set $f=1$, and drop other terms with negative powers of $\rho$, so our wave equations \eqref{3.7} simplify; for both the tensor and vector modes the solution becomes just a Hankel function:
\beq\label{Hankel1}
\psi_\infty=\sqrt{r} H_{n\omega_c}^{(1)}\left(r\sqrt{\omega^2-m^2}\right), \qquad \omega_c=\frac{l}{n}+\frac{1}{2}.
\eeq
Here, we have chosen to use $H^{(1)}$ because it is outgoing near $r\rightarrow\infty$, provided that $\omega^2>m^2$.  Following the results of \cite{Keeler:2014hba}, we will take the boundary condition appropriate to the physical value of $m^2$, in this case $m=0$, and simply analytically continue it to all $m^2$.  Consequently, the boundary condition at $r=\infty$ is really that we should have none of the $H^{(2)}$ function present there.

In the sections below, we will proceed by a matched expansion.  That is, we will solve the wave equation in the near-horizon region first, apply the horizon boundary condition, and then match this solution to the solution $\psi_\infty$ from the far zone by using the overlap region.  The requirement to have no $H^{(2)}$ behavior by the time we reach the far zone will result in a discrete set of allowed values for $\omega$.

\subsection{$SO(n+2)$ Vector modes}
Since we want to concentrate on the near-horizon physics, we will first consider the quasinormal modes that \cite{Emparan:2014aba} found to have nontrivial dependence on the near-horizon region:  these are their decoupled modes.  For vectors, these modes have both frequency and angular momentum satisfying $l,\omega=\calO(1)$ in a $1/n$ expansion.  As we will see in Section \ref{DecoupledVector}, these modes can indeed only be obtained by setting $\mu_0=1/2$, or $\Delta=\calO(1)$, as we discussed in Section \ref{3.1}.

After finding the $\omega$ for these decoupled modes, we then proceed to argue that indeed the choice $\Delta=\calO(1)$ eliminates all of the modes with $\omega$ or $l$ at size $n$ or larger; that is, it eliminates all of the modes which have support in the far zone, the ones which are not decoupled from the asymptotics in which the black hole is placed.  Consequently, in these two sections, we show that our choice of $\Delta$ combined with the restriction $\Delta=\calO(1)$, effectively studies only the physics of the near-horizon, decoupled modes.

\subsubsection{Decoupled Vector modes}\label{DecoupledVector}
The decoupled modes correspond to $l,\omega=\calO(1)$. As found in \cite{Emparan:2014aba}, these modes are necessarily normalizable within the near-horizon region.  They evade having any $H^{(2)}$ behavior at $r=\infty$ by being exponentially suppressed there.  In other words, decoupled modes have their primary support within the near-horizon region, and are thus normalizable in the asymptotic range of the near-horizon region. 

Following \cite{Emparan:2014aba}, we study the wave equation \eqref{3.7} perturbatively, order by order in the $1/n$ expansion. In order to perform the expansion, we expand all quantities in powers of $1/n$:
\begin{equation}\label{20} 
\psi=\sum\limits_{k\geq0} \frac{\psi_V^{(k)}}{n^k},\qquad
 \mathcal{L}=\sum\limits_{k\geq0} \frac{\mathcal{L}_V^{(k)}}{n^k},\qquad
  \omega=\sum\limits_{k\geq0}\frac{\omega_{(k)}}{n^k},\qquad
  U_V =\sum_{k\geq0}\frac{U_V^{(k)}}{n^k}.
\end{equation}
We additionally expand $\Delta$,  using the definition of $\mu$ from \eqref{3.5}:
\beq
  \Delta=n\left(\mu_0-\frac{1}{2}\right)+\sum_{k\geq0}\frac{\mu_{(k+1)}}{n^k}.
\eeq
Note we have left $\mu_0$ arbitrary rather than setting it to $1/2$; we do so in order show that the only quasinormal modes with $l, \omega =\calO(1)$ require $\mu_0=1/2$.  Since $\mu=1/2$ ensures $\Delta=\calO(1)$,  this result will support our choice of $\Delta$ as the meromorphic parameter.

To leading order in $1/n$, \eqref{3.7} reduces to
\begin{equation}\label{3.8}
\left(\mathcal{L}^{(0)}+U_V^{(0)}+\frac{(\rho-1)(4\mu_{0}^2-1)}{4 \rho}\right)\psi_V^{(0)}=0,
\end{equation}
with
\begin{equation}
\mathcal{L}^{(0)}=-(\rho-1)\frac{d}{d\rho}\left((\rho-1)\frac{d}{d\rho}\right),
\end{equation}
and
\begin{equation}
U_V^{(0)}=\frac{(\rho-1)(\rho-3)}{4\rho^2}.
\end{equation}

Using $\omega=\calO(1)$, the horizon boundary condition \eqref{quasiBC} can also be written order by order:
 \begin{align}
 \label{23}
 \psi^{(0)} &\underset{\rho\rightarrow 1}{\propto}  1,
\\
\label{24}
 \psi^{(1)}&\underset{\rho\rightarrow 1}{\propto} -i \omega_{(0)}\ln(\rho-1),
\\
\label{25}
 \psi^{(2)}&\underset{\rho\rightarrow 1}{\propto} -i \omega_{(1)}\ln(\rho-1)-\frac{1}{2}\omega^2_{(0)}(\ln(\rho-1))^2,
 \end{align}
 where we have fixed the overall amplitude of $\psi$ via \eqref{23} for convenience.  As highlighted above, at the boundary of the near-horizon region $\rho>>1$, we will insist on normalizability.

The solution to the lowest order equation \eqref{3.8} and the boundary condition \eqref{23} is
\begin{equation}\label{29c}
\psi_V^{(0)}=\rho^{3/2}{}_2F_1\left(\frac{3}{2}+\mu_{0},\frac{3}{2}-\mu_{0};1;1-\rho\right).
\end{equation}
To get the behavior in the overlap region, we expand the remaining piece around $\rho\rightarrow\infty$:
\begin{align}\label{42}
\psi_V^{(0)}=&\rho^{-\mu_{0}} 
\frac{\Gamma{(1)}\Gamma{(-2\mu_{0})}}{\Gamma{(3/2-\mu_{0})}\Gamma{(-1/2-\mu_{0})}}{}{_2}
F_1\left(3/2+\mu_{0},-1/2+\mu_{0},2\mu_{0}+1;1/\rho\right)
\\\nn
&+\rho^{\mu_{0}}\frac{\Gamma{(1)}\Gamma{(2\mu_{0})}}{\Gamma{(3/2+\mu_{0})}\Gamma{(-1/2+\mu_{0})}}{}{_2}F_1\left(3/2-\mu_{0},-1/2-\mu_{0},-2\mu_{0}+1;1/\rho\right).
\end{align}
From this form we can easily see that for $\mu>0$, the second line will be non-normalizable.  We also find
\begin{equation}\label{NormBC}
\psi(\rho\rightarrow\infty) \rightarrow \rho^{-\mu_{0}},
\end{equation}
that is, any $\rho^{\mu_{0}}$ behavior should be absent.  We use the analytic continuation of this boundary condition when we consider negative or imaginary values for $\mu_0$.

Applying the boundary condition \eqref{NormBC} to the solution \eqref{42} requires that one of the $\Gamma$ functions in the denominator of the second line have a pole, or in other words
\begin{equation}\label{43}
\mu_{0}=-k+\frac{1}{2} \qquad \text{for }k\in\mathbb{Z} \text{ and }k\geq 0.
\end{equation}
Since the values of $\mu_0$ that can possibly solve our boundary condition turn out to be half-integers, we actually need to revaluate the expansion in \eqref{42}.  When the first two parameters in a hypergeometric function differ by an integer, in this case $2\mu_0$, then the expansion takes a more complicated form.  We first rewrite the function around $\rho=1$ in terms of $k$:
\beq\label{kHypergeo}
\psi_V^{(0)}=\rho^{3/2}{}_2F_1 \left(2-k,k+1;1;1-\rho\right).
\eeq
For both $k=0$ and $k=1$, this solution becomes exactly
\begin{equation}\label{44}
\left.\psi_V^{(0)}\right\rvert_{k=0,1}=\rho^{-1/2}.
\end{equation}
However, if we rewrite this solution in terms of $\mu_0$, we find
\begin{equation}
\left.\psi_V^{(0)}\right\rvert_{k=0}=\rho^{-\mu_0}, \qquad \left.\psi_V^{(0)}\right\rvert_{k=1}=\rho^{\mu_0}.
\end{equation}
Now we can see that only the first option, $k=0$, satisfies the normalizability condition \eqref{NormBC}.  If we can also rule out $k\geq2$, then we will have $\mu_0=1/2$, or $\Delta=\calO(1)$ in $1/n$, as we wanted.

Before we analyze the $k=0$ case in more detail, let us examine these $k\geq2$ solutions.  Although the expansion \eqref{42} is no longer valid, we can see from the rewriting \eqref{kHypergeo} that for all integers $k\geq2$, the solution \eqref{29c} takes the exact form
\begin{align}\label{29d}
\psi_V^{(0)}&=\rho^{3/2}(a_{k-2} \rho^{k-2}+a_{k-3}\rho^{k-3}+...+ a_1 \rho+ a_0 )
\\
&=(a_{k-2} \rho^{-\mu_0}+a_{k-3}\rho^{-\mu_0-1}+...+ a_1 \rho^{5/2}+ a_0 \rho^{3/2}).
\end{align}
Since the series always terminates with $a_0$, no $\rho^\mu_0$ behavior is present, so these solutions do satisy the boundary condition at infinity \eqref{NormBC}.  However, as we will show in Appendix \ref{B}, continuing the perturbation procedure to one higher order results in quasinormal modes $\omega$ which do not satisfy the Euclidean regularity condition \eqref{QNMcondition} for any value of $\mu_1$. Accordingly, we will not consider $k\geq2$ further in the main body of the paper.

Since the only mode remaining is $k=0$ (corresponding to $\mu_0=1/2$ or $\Delta = \calO(1)$ in $1/n$), we now find its quasinormal mode frequencies.   Using the $\calO(1)$ solution \eqref{44} and expanding to next order in $1/n$,  the wave equation \eqref{3.7} becomes
\begin{equation}\label{50}
4\rho^{5/2}(\rho-1)\frac{d^2\psi^{(1)}}{d\rho^2}+4\rho^{5/2}\frac{d\psi^{(1)}}{d\rho}-\sqrt{\rho}(\rho-3)\psi^{(1)}+2\rho(1-2l-2\mu_1)+4=0.
\end{equation}
The general solution for $\psi^{(1)}$ for the physical region $\rho>1$ becomes 
\beq
\psi^{(1)}(\rho)=C_{1/2}\sqrt{\rho}+\frac{2 C_{-1/2}-\log \rho}{2\sqrt{\rho}}+\left(1-l-\mu_1+C_{1/2}\right)\frac{\log (\rho-1)}{\sqrt{\rho}},
\eeq
where $C_{1/2}$ and $C_{-1/2}$ are arbitrary constants.  Since the $C_{1/2}\sqrt{\rho}$ term violates the normalizability condition \eqref{NormBC}, we set $C_{1/2}=0$. Expanding the solution around $\rho=1$ in order to apply the horizon boundary condition \eqref{24}, we obtain
\beq
\psi^{(1)}(\rho)\underset{\rho\rightarrow 1}{\propto} (1-l-\mu_1)\log(\rho-1)+C_{-1/2} + \calO(\rho-1)^1.
\eeq
Matching to \eqref{23} requires $C_{-1/2}=0$,  while matching to \eqref{24} gives the leading order quasinormal frequency
\begin{equation}\label{51}
\omega_{(0)}=-i(l+\mu_1-1).
\end{equation}
We have now found the leading-order term in the quasinormal mode frequencies under the assumption that $\omega, \, l =\calO(1)$ in $1/n$, and that the modes we are interested in should be normalizable in the near zone.  Normalizability in the near zone ensures that these modes capture information about the dynamics of the near-horizon, membrane region, and additionally justifies our choice to concentrate on $\Delta =\calO(1)$.

\subsubsection{Non-decoupled vector modes}\label{NonDecoupVector}
As we will now show, if we allow either (or both) of $\omega$ and $l$ to be of order $\calO(n)$ or larger, the resulting modes are non-normalizable in the near-horizon geometry.

Although we have justified the choice of the mass parameter $\Delta$ and the condition $\Delta=\calO(1)$, we will begin by allowing a $\Delta$ to grow at order $n^2$:
\beq
\Delta=n(\mu-1/2)=\mu_{-1} n^2 + n(\mu-1/2)+\mu_1+\mu_2/n+\calO(1/n^2).
\eeq
Next, we assume both $l,\, \omega$ are no larger than $\calO( n^2)$:
\begin{align}
l&=l_{-2}n^2+l_{-1}n+l_0+l_1/n+\calO(1/n^2),
\\
\omega&=\omega_{-2}n^2+\omega_{-1}n+\omega_0+\omega_1/n+\calO(1/n^2).
\end{align}
This assumption is just to simplify the argument; as we will show, both $l_{-2}$ and $\omega_{-2}$ (as well as $\mu_{-1}$ are forced to be zero when solving the vector version of \eqref{3.7}, and any larger terms would be as well.
With these assumptions, the vector wave equation \eqref{3.7} at order $n^2$ is
\beq\label{minus2vanish}
\left(l_{-2}^2-\frac{\omega_{-2}^2}{1-1/\rho}+\mu_{-1}^2\right)\psi^{(0)}=0.
\eeq
A nonzero solution is only possible if $\omega_{-2}=0$ and $l_{-2}^2+\mu_{-1}^2=0$.
Next, we examine the equation \eqref{3.7} at order $n$, imposing $\omega_{-2}=0$; we find
\beq
\left(l_{-2}(1+2l_{-1})+2\mu_{-1}\mu_0-2l_{-2}^2 \log \rho\right)\psi^{(0)}=0
\eeq
Since the $\log \rho$ term is independent of the others, its coefficient must vanish; thus, we find $l_{-2}=0$ which implies $\mu_{-1}=0$ from the order $n^2$ condition.  Consequently, we have $\mu_{-1}=l_{-2}=\omega_{-2}=0$ from the $\calO(n^2)$ and $\calO(n)$ equations combined.

Now, using our intuition from Section \ref{DecoupledVector} above, we expect $\mu_0=1/2$, so that the first nonzero term in $\Delta$ is $\m_1$ at order $n^0$.
Thus in the near-horizon coordinates at leading order $\calO(n^0)$, \eqref{3.7} becomes
\begin{equation}
-(\rho-1)\frac{d^2\psi^{(0)}}{d\rho^2}-\frac{d\psi^{(0)}}{d\rho}-\frac{\hatomega^2}{\rho-1}\psi^{(0)}+\frac{4\omegac^2\rho-3}{4\rho^2}\psi^{(0)}=0,
\end{equation}
where $\omegac$ is the $\calO(n^0)$ component of $\omega_c$ from the far region solution \eqref{Hankel1}, so $\omegac=1/2+l_{-1}$.

The solution to the above equation satisfying the horizon boundary condition \eqref{quasiBC} is
\begin{equation}\label{101}
\psi^{(0)}=(\rho-1)^{-i \hatomega}\rho^{3/2}\hspace{1mm}{_2}F_1(1+q_{+},1+q_{-},q_++q_-;1-\rho),
\end{equation}
where 
\begin{equation}\label{qplusminus}
q_{\pm}=\frac{1}{2}-i\hatomega \pm \sqrt{\omegac^2-\hatomega^2}.
\end{equation}
In the overlap zone we have large $\rho$, so the solution \eqref{101} becomes %
%
%
\begin{align}\label{3.28}
\psi^{(0)}\underset{\rho\rightarrow\infty}{\sim}\rho^{\frac{q_- -q_+}{2}}\left(\frac{
\Gamma\left(q_++q_-\right)
\Gamma\left(q_--q_+\right)
}{
\Gamma\left(q_--1\right)
\Gamma\left(q_-+1\right)
}\right)+
\rho^{\frac{q_+ -q_-}{2}}\left(\frac{
\Gamma\left(q_++q_-\right)
\Gamma\left(q_+-q_-\right)
}{
\Gamma\left(q_+-1\right)
\Gamma\left(q_++1\right)
}\right).
\end{align}
As we can see, these modes have both normalizable and non-normalizable behaviors in the asymptotic regime of the near-horizon region, that is, at large $\rho$. In order to remove the nonnormalizable behavior (which could be either term depending on the parameters), we would need either $q_+$ or $q_-$ to solve
\beq
q_\pm-1=-k, \qquad k\in\mathbb{Z} \text{ and } k\geq 0.
\eeq
Rewriting as a condition on $\hatomega$, we have
\begin{align}
i\hatomega&=\frac{k^2-k-l_{-1}^2-l_{-1}}{2k-1}
\nn\\\label{vectoromegaminus1}
&=-p, \qquad \text{where }p,\, k, \, l_{-1} \in \mathbb{Z}_{\geq0}.
\end{align}
In the last line, we have added the requirement to additionally satisfy the regularity condition \eqref{QNMcondition}, for a black hole with 
$r_0=1$ or $2\pi T=1$.  The only possible solution with integers is
\beq
p=\hatomega=0, \quad k=l_{-1}+1.
\eeq
However, when $\hatomega=0$, then $\omega=\calO(1)$ and we have
\beq
q_+=1+l_{-1},\quad q_-=-l_{-1}.
\eeq
Thus the expansion \eqref{3.28} fails to be valid as $(q_+-q_{-})$ is an integer, but \eqref{101} has a terminating expansion of the form
\begin{equation}\label{3.29}
\rho^{3/2}(a_1+a_2\rho+....+a_{l_{-1}}\rho^{l_{-1}-1})
\end{equation}
We see that these modes have only non-normalizable behavior at  large $\rho$, and thus should not be included.  Consequently, we find that there are no poles with $\Delta = \calO(1)$ due to vector modes with either $\omega = \calO(n)$ or larger, or with $l=\calO(n)$ or larger. Non-decoupled modes do not contribute poles in the $\Delta=\calO(1)$ region. We will return to finding the poles associated with the decoupled vector modes in Section \ref{sec:QNMresults} below, after providing a brief argument that no tensor modes contribute.

\subsection{$SO(n+2)$ Tensor modes}\label{NoTensor}
In the tensor case \cite{Emparan:2014aba} found only non-decoupled modes, that is only modes with large profile far from the horizon.  Given our vector result (that no non-decoupled modes contribute relevant poles), we expect a similar result for tensors.  The argument is quite similar to the vector case.

Since the only difference between the vector potential \eqref{2.8} and the tensor potential \eqref{2.7} first shows up at $\calO(1)$ in $1/n$, the arguments around \eqref{minus2vanish} leading to the result $\omega_{-2}=l_{-2}=0$ apply to tensors as well as vectors, again assuming $\mu_0=1/2$, by our definition.

Consequently, the tensor equation \eqref{3.7} becomes, at $\calO(1)$ in $1/n$,
\begin{equation}
-(\rho-1)\frac{d^2\psi^{(0)}}{d\rho^2}-\frac{d\psi^{(0)}}{d\rho}-\frac{\hatomegaT^2}{\rho-1}\psi^{(0)}+\frac{4\omegacT^2\rho+1}{4\rho^2}\psi^{(0)}=0,
\end{equation}
where again $\omega_{-1}$ is the $\calO(n)$ piece in $\omega$, and $\omegacT=1/2+l_{-1}$ with $l_{-1}$ the $\calO(n)$ piece in the angular momentum $l$, as in the vector case.

The solution to the above equation is 
\begin{equation}\label{3.33}
\psi^{(0)}=(\rho-1)^{-i \hatomegaT}\rho^{1/2}\hspace{1mm}{_2}F_1(q_{+},q_{-},q_++q_-;1-\rho),
\end{equation}
where $q_\pm$ are as in \eqref{qplusminus}.
In the overlap zone at large $\rho$, \eqref{3.33} expands as

%

%
\begin{equation}\label{3.36}
\psi^{(0)}\underset{\rho\rightarrow\infty}{\sim}\rho^{\frac{q_- -q_+}{2}}\frac{
\Gamma\left(q_++q_-\right)
\Gamma\left(q_--q_+\right)
}{
\left(\Gamma\left(q_-\right)\right)^2
}+
\rho^{\frac{q_+ -q_-}{2}}\frac{
\Gamma\left(q_++q_-\right)
\Gamma\left(q_+-q_-\right)
}{
\left(\Gamma\left(q_+\right)\right)^2
}.
\end{equation}
For generic values of the parameters, we see these modes have both normalizable and non-normalizable behaviors in the asymptotic regime.  
As before, we might guess that $q_\pm=-k$ for a nonnegative integer $k$ might allow us to remove the non-normalizable behavior.  Quite similarly to \eqref{vectoromegaminus1}, upon adding the requirement to satisfy the regularity condition \eqref{QNMcondition}, we find 
\beq
i\omega_{-1}=-p=\frac{k^2+k-l_{-1}^2-l_{-1}}{2k+1}.
\eeq
Again, there is only one possible solution with integers:
\beq
p=\omega_{-1}=0, \quad k=l_{-1}.
\eeq
As before, showing $\omega_{-1}=0$ is equivalent to requiring $\omega=\calO(1)$ in $1/n$; the expansion in \eqref{3.36} is not valid here because $q_+-q_-$ is an integer, but fortunately the solution \eqref{3.33} has a terminating expansion, of the form
\beq
\psi^{(0)}=\rho^{3/2}\left(a_0+a_1\rho+\ldots+a_{l_{-1}}\rho^{l_{-1}}\right).
\eeq
As in the vector case this solution has only non-normalizable behavior at large $\rho$, so it should not be included. We have now eliminated any solution with either $l=\calO(n)$ or greater, or $\omega=\calO(n)$ or greater.

When $\omega$ and $l$ both are $\calO(1)$, the leading order tensor equation \eqref{3.7}  becomes
\begin{equation}
(\rho-1)\frac{d^2\psi^{(0)}}{d\rho^2}+\frac{d\psi^{(0)}}{d\rho}-\frac{\rho+1}{4\rho^2}\psi^{(0)}=0,
\end{equation}
and its general solution is
\begin{equation}\label{3.38}
\psi^{(0)}=C_1\sqrt{\rho} +C_2\sqrt{\rho}\ln(1-\rho^{-1}).
\end{equation}
This mode is also not allowed, because regardless of the subleading values of $\Delta$, there is no way to satisfy the normalization \eqref{NormBC} and horizon \eqref{24} boundary conditions.

Accordingly, as per our expectation, no tensor modes contribute poles in the region $\Delta=\calO(1)$.

\subsection{Quasinormal mode results}
\label{sec:QNMresults}
In sections \ref{NonDecoupVector} and \ref{NoTensor}
 we showed that there are no allowed tensor and non-decoupled vector modes, so the only poles we are interested in are the ones corresponding to decoupled vector modes.

In order to find the poles, we will need the Euclidean regularity condition \eqref{QNMcondition}.  Since we have set $r_0=1$, the black hole temperature becomes $2\pi T=1$, and we have
\begin{equation}\label{3.19}
\omega_{(0)}=2\pi i pT=ip.
\end{equation}
%


Requiring that the decoupled vector mode frequencies \eqref{51} satisfy \eqref{3.19} gives
\begin{equation}\label{52a}
\mu_1=1-l-p.
\end{equation}
Thus the corresponding poles occur at
\begin{equation}\label{QNMpoles}
\Delta^*_V=n\left(\mu_0-\frac{1}{2}\right)+\mu_1+\calO(1/n)=1-l-p+ \mathcal{O}(1/n),\quad p, l\in\mathbb{Z}_{\geq 0},
\end{equation}
where we have used $\mu_0=1/2$. 
For anti-quasinormal modes, the poles can be obtained in a similar way.  The boundary condition at the horizon has a sign change, as in \eqref{antiquasiBC},  which means the first order boundary condition \eqref{24} should become instead
\beq
\bar{\psi}^{(1)} \underset{\rho\rightarrow1}{\propto} i\bar{\omega}_{(0)}\ln (\rho-1),
\eeq
while the regularity condition \eqref{QNMcondition} becomes $\bar{\omega}_{(0)}=i p$, where $p\in\mathbb{Z}_{<0}$. We find that the poles for anti-quasinormal modes instead occur at 
\begin{equation}\label{antiQNMpoles}
\bar{\Delta}^*_V=1-l+p+ \mathcal{O}(1/n),\quad p\in\mathbb{Z}_{<0}.
\end{equation}
We note that the $p=0$ case is not allowed here as it is already accounted for in  \eqref{QNMpoles}. We summarize our results in Table \ref{QNMtable}.

\begin{table}
\begin{center}
 \begin{tabular}{||c |c| c| c||} 
 \hline
  & Poles & p &l \\ [0.5ex] 
 \hline\hline
 QNM & $\Delta^*_V=1-l-p+ \mathcal{O}(1/n)$ &$\in\mathbb{Z}_{\geq 0}$ &$\in\mathbb{Z}_{\geq 0}$  \\ 
 \hline
 Anti-QNM & $\bar{\Delta}^*_V=1-l+p+ \mathcal{O}(1/n)$ & $\in\mathbb{Z}_{< 0}$&$\in\mathbb{Z}_{\geq 0}$ \\
 \hline
\end{tabular}
\caption{Summary: Poles for vector (anti)quasinormal modes in the $\Delta$ complex plane.} \label{QNMtable}
\end{center}
\end{table}
For both the QNM and anti-QNM series, additional care should be taken for mode numbers less than the spin of the particle being considered; since we are considering gravitons, we should reanalyze the solutions with $l=0,1$ and $p=0,1$, to ensure the modes are actually integrable at the origin of the Euclidean thermal space.  However, since any necessary adjustment would only result in removing a finite number of poles, and we will concern ourselves below mainly with the behavior of the one-loop determinant in the $\Delta\rightarrow \infty$ limit, we postpone this analysis to future work.  For more details on this issue, the interested reader may consult Append B.3 of \cite{Castro:2017mfj}.
\section{Writing the One-loop determinant}\label{4}
As we will show, it is possible to express the $SO(n+2)$ vector and tensor portions of the graviton one-loop determinant in the large-dimension Schwarzschild background directly in terms of the Hurwitz zeta function and its derivatives. Since the $SO(n+2)$ tensor does not contribute any poles in the $\Delta =\calO(1)$ regime, our work in this section primarily relies on the poles due to vector quasinormal modes \eqref{QNMpoles} and anti-quasinormal modes \eqref{antiQNMpoles}.  We defer treating the $SO(n+2)$-scalar graviton contribution for future work, but make some commentary in appendix \ref{C}.

\subsection{Expressing $Z_V$ in terms of Hurwitz $\zeta$}
Using Weierstrass's factorization theorem as reviewed in Section \ref{sec:reviewQNM} the one-loop determinant in the large $n$ limit becomes
\begin{equation}
Z_V=e^{\rm{Pol(\Delta_V)|_{n\rightarrow \infty}}} \prod \limits_{l\geq0,p\geq0}\left(\Delta_V-(1-l-p)\right)^{-D_l}\prod \limits_{l\geq0,p<0}\left(\Delta_V-(1-l+p)\right)^{-D_l},
\end{equation}
where the degeneracy $D_l$ of each frequency equals the degeneracy of the $l\rm{th}$ angular momentum eigenvalue on $S^{n+1}$
\begin{equation}
D_l^{n+1}=\frac{2l+n}{n}{l+n-1 \choose n-1}.
\end{equation}
Taking the logarithm and rearranging, we find
\begin{align}
-\log Z_V+\mathrm{Pol}(\Delta_V)&=2\sum\limits_{l\geq0,p\geq0}D_l \log\left(\Delta_V-(1-l-p)\right)-\sum\limits_{l\geq0}D_l \log\left(\Delta_V-(1-l)\right)
\nn\\
&= 2\sum_{j\geq0}\left(\sum_{l=0}^j D_l\right)\log(\Delta_V +j-1)-\sum_{j\geq0} D_j \log (\Delta_V+j-1)
\nn\\
&=\sum_{j\geq0}\left(-1+2\sum_{l=0}^j D_l\right)\log(\Delta_V +j-1)
\nn\\
&=\sum\limits_{j\geq0}\tilde D(j)\log\left(\Delta_V+j-1\right),
\label{57a}
\end{align}
where $\tilde D(j)$ is given by 
\begin{equation}\label{EffDegen}
\tilde D (j)=\frac{n+(2j+n)^2}{n(n+1)}{ j+n-1\choose n-1}.
\end{equation}
We now follow \cite{Denef:2009kn} to rewrite this sum in terms of the Hurwitz zeta function $\zeta(s,x)$, defined as
\beq
\zeta(s,x) = \sum_{q=0}^\infty\frac{1}{(x+q)^s}.
\eeq
We then use the derivative $\zeta'(s,x)=\frac{\partial}{\partial s}\zeta(s,x)$ and the shift operator $\delta_s$ which acts on a function $f(s)$ as
\begin{equation}
\delta_s f(s)\equiv f(s-1),
\end{equation}
to rewrite \eqref{57a} in a compact form. The rewriting will rely on the relation
\beq
\sum_{q=0}^{\infty}\frac{\log(q+x)\text{Poly}(q)}{(q+x)^s}=-\text{Poly}(-x+\delta_s) \zeta'(s,x),
\eeq
which is accurate for any polynomial $\text{Poly}(q)$.  We find
\beq\label{ZVinzeta}
\log Z_V
=\text{Pol}(\Delta_V)-\log(\Delta_V-1)+\left.\tilde{D}\left(1-\Delta_V+\delta_s\right) \zeta'(s,\Delta_V)\right\rvert_{s=0}.
\eeq
%
%
\subsection{Matching with the heat kernel expression}\label{4.2}
In this section we compute the large $\Delta$ limit of $\log Z_V$. We will then compare these results to the heat kernel curvature expansion, at least for the largest terms.  In a successful comparison, this procedure fixes the polynomial piece $\text{Pol}(\Delta_V)$, as done in \cite{Denef:2009kn}.  In our case, we do find a match between the non-polynomial terms in the heat kernel expansion appropriate to $D-1$ dimensions, and the non-polynomial terms in the quasinormal mode method, indicating that a higher order calculation should be able to fix the polynomial piece. We discuss our interpretation of this result, as well as some caveats, in this section and in the discussion in Section \ref{5}.
\subsubsection{Large $\Delta$ limit using QNM method}
To get the leading large $\Delta$ behavior of \eqref{ZVinzeta}, 
 we begin by finding the highest power of $j$ in the degeneracy polynomial \eqref{EffDegen}:
\begin{equation}
\tilde{D}(j)=\frac{4}{(n+1)!}j^{n+1}+\#\, j^n+\ldots,
\end{equation}
where $\#$ refers to a coefficient we will not need below. We will also use the expansion of the Hurwitz zeta derivative at large argument:
 \begin{equation}\label{62a}
 \zeta'(s,\Delta_V)=\sum\limits_{k=0}^\infty \frac{(-1)^k B_k}{k!}\left[ \frac{\Gamma(k+s-1)}{\Gamma (s)}(-\log \Delta)+\partial_s\left(\frac{\Gamma(k+s-1)}{\Gamma (s)}\right)\right]\Delta^{1-s-k},
 \end{equation}
where $B_k$ are the Bernoulli numbers. 
Using these expressions, we expand the final term in \eqref{ZVinzeta} at large $\Delta_V$, keeping only the leading terms.  We find
\begin{align}
\nn
\left.\tilde{D}\left(1-\Delta_V+\delta_s\right) \zeta'(s,\Delta_V)\right\rvert_{s=0} &= \frac{4}{(n+1)!}\left.(1-\Delta_V+\delta_s)^{n+1}\zeta'(s,\Delta_V)\right\rvert_{s=0}+\ldots
\\
&= \frac{4}{(n+1)!}\sum_{r=0}^{n+1} {n+1 \choose r} (1-\Delta_V)^{n+1-r}\left.(\delta_s)^r\zeta'(s,\Delta_V)\right\rvert_{s=0}+\ldots
\end{align} 
In order to find the highest power of $\Delta_V$ in this expression, we need to find the largest power of $\Delta_V$ in
\begin{align}
\left.(\delta_s)^r \zeta'(s,\Delta_V)\right\rvert_{s=0}&=\zeta'(-r,\Delta_V)\nn
\\\nn
&=B_0\left[\frac{\Gamma(-1-r)}{\Gamma(-r)}(-\log \Delta_V)+\left.\partial_s\left(\frac{\Gamma(s-1)}{\Gamma(s)}\right)\right\rvert_{s=-r}\right]\Delta^{r+1}+ \#\, \Delta^{r}+ \ldots
\\
&=\left[\frac{1}{1+r}\log \Delta_V-\frac{1}{(1+r)^2}\right]\Delta^{r+1}+\# \Delta^r+\ldots
\end{align}
We then find the leading terms at large $\Delta_V$ are
\begin{align}\label{LargeDeltaQNM}
\left.\tilde{D}\left(1-\Delta_V+\delta_s\right) \zeta'(s,\Delta_V)\right\rvert_{s=0} \xrightarrow{\Delta\rightarrow\infty}
& \sum_{r=0}^{n+1} \frac{4(-1)^{n+1-r}}{(r+1)!(n+1-r)!}\left[\log\Delta_V-\frac{1}{1+r}\right]\Delta^{n+2}.
\end{align}
Following \eqref{ZVinzeta}, these terms are also the largest in $\log Z_V-\text{Pol}(\Delta_V)$ as well.  In particular the logarithmic terms cannot be cancelled by the polynomial piece, so the leading behavior of the determinant satisfies
\beq\label{QNMlargeDelta}
\log Z_V \underset{\Delta\rightarrow\infty}{\propto} \Delta_V^{n+2}\log \Delta_V  = \Delta_V^{D-1}\log \Delta_V .
\eeq

\subsubsection{Heat kernel calculation}
We will now find the large $\Delta$ behavior for the full $\log Z$, via the heat kernel curvature expansion (for a review, see e.g. \cite{Vassilevich:2003xt}). In this expansion, we expand the determinant at large mass as
\begin{equation}\label{4.10}
\log Z=\mathrm{const}-(4\pi)^{-\D/2}\sum\limits_{q=0}^{\D} a_q\int_\epsilon^{\infty}\frac{dt}{t} t^{\frac{q-\D}{2}}e^{-t m^2}+\mathcal{O}(1/m),
\end{equation} 
where $\D$ denotes the effective total dimension of the spacetime, $a_q$ denotes the heat kernel curvature coefficients and $\epsilon$ is a regulator which we will take to zero below. The heat kernel curvature coefficients are zero for odd $q's$ for spacetimes without a boundary; as we will show below, the $a_0$ term gives the leading large $\Delta$ behavior, and is present regardless of the dimension.

We find the leading large $m$ behavior for the $q$th term in \eqref{4.10} by evaluating the integral and expanding for small regulator $\epsilon$.  For odd dimensions, the leading term goes like $m^{\D-q}$, or in terms of $\Delta$ it goes like $\Delta^{\D-q}$:
\begin{equation}\label{4.11odd}
\log Z\underset{m\rightarrow\infty}{\propto}m^{\D-q}.
\end{equation} 
For even $\D$, we find instead
\begin{equation}\label{4.11even}
\log Z\underset{m\rightarrow\infty}=c_1m^{\D-q}+c_2 m^{\D-q}\log\left(m^2\epsilon\right),
\end{equation} 
where $c_1$ and $c_2$ are constants dependent on the $a_n$.
In both cases, the largest power of $m$, and thus of $\Delta$, corresponds to the term with $q=0$. 
For $q=0$, \eqref{4.11even} explicitly becomes
\begin{equation}\label{evenD_q=0}
\log Z\underset{m\rightarrow\infty}=-a_0\left(-\frac{1}{4\pi}\right)^{\D/2}\left[\frac{H(\D/2)-\gamma-\log(m^2 \epsilon)}{(\D/2)!}\right]m^{\D},
\end{equation}
where $H(x)$ is the $x$th harmonic number. Expanding \eqref{evenD_q=0} to leading order in terms of large $\Delta$ where $\Delta$ solves \eqref{52c} gives
\begin{equation}\label{LargeDeltaHeat}
\log Z\underset{\Delta\rightarrow\infty}=-a_0\left(-\frac{1}{4\pi}\right)^{\D/2}\left[\frac{H(\D/2)-\gamma-2\log\Delta-\log\epsilon}{(\D/2)!}\right]\Delta^{\D}.
\end{equation}
Since $a_0=\int d^{\D}x\sqrt{g_0}$, the leading term is thus proportional to the (regulated) volume.  In terms of $\Delta$, the leading behavior of this volume-dependent term becomes $\Delta^{\D}$ for odd $\D$, and $\Delta^{\D} \log \Delta$ for even $\D$. We also note that no $\log \Delta$ terms are present in the odd $\D$ expansion, regardless of how many terms we consider.

Now, we want to compare $Z_V$ from the quasinormal mode method \eqref{ZVinzeta} to the heat kernel result.  Expanding \eqref{ZVinzeta} at large $\Delta$ (using \eqref{LargeDeltaQNM}), we find
\beq\label{fullQNMlargeDelta}
\log Z_V
\underset{\Delta_V\rightarrow\infty}=
\text{Pol}(\Delta_V)+ \sum_{r=0}^{D-2} \frac{4(-1)^{D-2-r}}{(r+1)!(D-2-r)!}\left[\log\Delta_V-\frac{1}{1+r}\right]\Delta_V^{D-1}.
\eeq
Comparing this quasinormal mode expression with the heat kernel results \eqref{4.11odd} and \eqref{4.11even}, we see that matching will only be possible when $\D$ is even.  

Comparing further with the large $\Delta$ expansion for even $\D$ as in \eqref{LargeDeltaHeat}, we find that the logarithmic behaviors in  \eqref{4.11even} and \eqref{fullQNMlargeDelta} match for even $\D$ when $\D=D-1$.  Although this result may at first be surprising, it shows that the heat kernel analysis should really be using an effective total dimension that is reduced by one, which reflects a clear picture of the membrane paradigm as mentioned in \cite{Bhattacharyya:2015dva}. This interpretation is quite strong since the $\log Z$ in \eqref{57a} is actually built out of poles corresponding to modes living in the membrane region, which is itself $D-1$ dimensional. 

Now, by matching the coefficient of $\log(\Delta_V)\Delta^{\D}$ in \eqref{fullQNMlargeDelta} and \eqref{LargeDeltaHeat}, using $\D=D-1$, we can calculate%
\footnote{Since we do not have a precise definition of the membrane region whose regulated volume $a_0$ should calculate, we cannot find it from first principles and must instead do a comparison. Since $a_0$ should be proportional to the (regulated) volume, we find this form encouraging; it is the ratio of sphere volumes for dimensions $D-1$ and $(D-1)/2$.  We do not have an interpretation of why this ratio should be the regulated volume appropriate to the membrane region.}
 $a_0$:
\begin{align}
a_0=\frac{-2(-4\pi)^{\frac{D-1}{2}}\left(\frac{D-1}{2}\right)!}{(D-1)!}.
\end{align}
Once we have obtained $a_0$, by equating the polynomial terms in \eqref{LargeDeltaHeat} and \eqref{fullQNMlargeDelta} at large $\Delta$ we can compute the leading behavior at large $\Delta_V$ of $\text{Pol}(\Delta_V)$:
\begin{equation}\label{PolVResult}
\text{Pol}_V(\Delta)= \frac{2}{\Gamma(D)}\left[H\left(\frac{D-1}{2}\right)-\gamma-\log\epsilon-2H(D-1)\right]\Delta^{D-1}.
\end{equation}
Thus, we claim the contribution to the partition function from the $SO(n+2)$ vector and tensor modes of the graviton in the large $D$ limit of a Schwarzschild background, for $D$ odd, is given by \eqref{ZVinzeta}, where the largest contribution to the polynomial is set by \eqref{PolVResult}.

There are a few caveats to this interpretation.  First, for even $D$ or odd $\D$, the heat kernel analysis and the quasinormal mode analysis disagree, since the quasinormal mode method indicates a logarithmic term whereas the heat kernel one only produces polynomials in $\Delta$.  We believe this behavior occurs because the large $D$ limit is not analytic when including both parities; since many behaviors depend on whether $D$ is odd or even, we should specify if we are taking a large $D$ limit for the odd case or the even case.  The presence of a logarithmic term in the quasinormal mode method indicates that the effective total heat kernel dimension $\D$ should be thought of as even, or the total spacetime dimension $D$ should be odd.

In addition to the concern about analyticity of the large $D$ limit, we should also mention an order of limits concern between large $\Delta$ and large $D$.  Recall that the quasinormal mode result in \eqref{57a} is calculated by taking the large $D$ limit; we then take the large $\Delta$ limit to compare with the heat kernel result.  Contrastingly, the heat kernel calculation first takes the large $\Delta$ limit for fixed $\D$.  

Another possible concern is that the heat kernel analysis may be expecting a fully gauge-fixed result, whereas we have only been able to analyze the gauge-independent tensor and vector variables in calculating \eqref{QNMlargeDelta}.  Accordingly, we have only claimed to calculate the polynomial piece due to the vector (and tensor) perturbations, and thus call our result in \eqref{PolVResult} $\text{Pol}_V$.  We expect a more complete calculation may have further contributions, particularly at subleading order in $\Delta$, to the polynomial piece due to scalar and gauge fixing perturbations.

Even given these caveats, we have shown that a consistent interpretation of our quasinormal mode calculation in the large dimension limit points to an effective total dimension of $D-1$ in the heat kernel analysis, which corresponds with a membrane-paradigm picture for the large dimension limit we analyze.

\section{Conclusion}\label{5}

We have studied the $SO(n+2)$ vector and tensor modes of the one-loop determinant, for fluctuations of the $n+3$-dimensional transverse traceless graviton, in a Schwarzschild black hole background in the large dimension limit.  We found that no tensor modes contribute poles to the one-loop determinant, while the vector modes result in poles at locations set by integers, specifically at
\beq
\Delta^*=1-l-p+\calO(1/n), \qquad p,l \in \mathbb{Z}_{\geq 0}.
\eeq
Using the locations of these poles, we constructed the expression \eqref{ZVinzeta} for the one loop determinant due to the vector modes, in terms of Hurwitz zeta functions.  Importantly, we found that the large $\Delta$ behavior of this determinant is $\Delta^{D-1}\log\Delta$, as in \eqref{QNMlargeDelta}.

We compared this result to that obtained from a preliminary heat kernel curvature expansion, presented in \eqref{4.11odd} for odd dimensions and \eqref{4.11even} for even dimensions; we find that the results match  when the heat kernel effective dimension $\D$ is even, and when $\D=D-1$ from the quasinormal mode analysis. As discussed in Section \ref{4.2}, this result reflects the dimensionality reduction in the leading power of $\Delta$ in $\log Z$ in agreement with the membrane paradigm picture of the large dimension black hole as presented in \cite{Bhattacharyya:2015dva}. 

We do believe the quasinormal mode result \eqref{QNMlargeDelta} as compared to the heat kernel result \eqref{4.11even} indicates a picture of effectively reduced dimensions in the large dimension limit.  However, there are a few caveats on our result.  First, there are two possible issues with the large dimension limit. As evidenced by the dependence on dimensional parity, the large dimension limit may not be analytic, or rather may not preserve the meromorphicity of the one-loop determinant. Additionally, there is a possible order of limits issue regarding taking the large $D$ vs. large $\Delta$ limits.  The quasinormal mode analysis relies on first taking a large $D$ limit, and then large $\Delta$; the heat kernel curvature expansion is itself already a large $\Delta$ limit, and we impose the large $D$ limit afterwards. 

Next, we do not perform the gauge fixing to obtain the full one-loop partition function for quasinormal modes.  The primary obstruction is that the generic gauge-fixing analysis of the partition function for the gauge-independent variables of \cite{Kodama:2003jz, Kodama:2000fa} has not been done.  Since their (and our) decomposition into scalar, vector, and tensor modes refers to the $SO(n+2)$ spherical symmetry of the spacetime, and not to the local Lorentz symmetry, we cannot use the \cite{Yasuda:1983hk} formula.  Accordingly, we leave this gauge-fixing problem to future work.

Next, our choice of mass parameter $\Delta =n(\mu-1/2)=-n/2+\sqrt{m^2+n^2/4}$ may appear somewhat arbitrary.  However, as we can see from the explicit behavior of the near- horizon modes in \eqref{42}, the appropriate boundary condition on $\psi$ in the asymptotia of the near-horizon region is indeed analytic in $\mu$ and thus in $\Delta$.  In fact this analyticity looks somewhat similar to that found for AdS in previous quasinormal mode studies (e.g \cite{Denef:2009kn, Keeler:2014hba, Keeler:2016wko,Castro:2017mfj};  we picked the name $\Delta$ for this parameter in deference to this analogy.

We should also justify the specific linear combination of the square root, that is, why we chose $\Delta=n(\mu-1/2)$ and not some other dependence on $n$.
As we showed in Section \ref{DecoupledVector}, this choice captures the physics of the near-horizon region at order $\Delta \sim \calO(n^0)$.  If we instead chose $\Delta/n$ as our mass parameter, then the order $\calO(n^0)$ poles would also include non-decoupled modes.  In other words, this choice of mass parameter would include the effect of modes which have nontrivial support in the asymptotic flat region.  Contrastingly, if we chose $n\Delta$, then all of the vector poles we found would also be pushed to infinite mass parameter; we could not resolve their effect even though these modes live entirely in the near-horizon region.
As we point out in Appendix \ref{C}, the scalar modes actually arise at $n^2(\mu-1/2)\sim \calO(n^0)$.  Accordingly, the choice of $n\Delta$ would capture the scalar modes; however it is not an acceptable mass parameter because it leaves out the near-horizon vector modes. 
The choice $\Delta=n(\mu-1/2)$ thus captures exactly the near-horizon physics for the tensor and vector modes that are our main focus,  at order $\calO(n^0)$.  We leave the study of scalar modes, which appear at $\Delta \sim \calO(1/n)$, to future work.

Better control over the large $D$ limit of the heat kernel curvature expansion (or perhaps of the heat kernel curvature expansion of the large $D$ limit) would elucidate the order of limits issues highlighted previously, and also aid in understanding correction terms present for finite $D$.  Our results are only appropriate in the strict $D\rightarrow\infty$ limit, but most physics of interest is at finite $D$. For a finite $D$, the dynamics of the asymptotic region may be important for the one loop determinant.  In the infinite $D$ limit, only the near-horizon dynamics matter. Since our analysis is only of the near-horizon dynamics, we lack a characterization of the correction terms for finite $D$. We leave such a characterization to future work.

We should also note that the similarity to the AdS analysis appears deeper than the choice of $\Delta$; the functional forms for $\psi$ at the pole locations $\Delta=\Delta^*$ as found in Section \ref{DecoupledVector} also appear quite similar to the AdS equivalents. Additionally, as pointed out in \cite{Emparan:2013xia}, the near-horizon region of the large $D$ limit is described by a two-dimensional string theory black hole geometry which has conformal symmetry. We hope to further explore this emergent symmetry in the future.

\section*{Acknowledgments}
The authors would like to thank Roberto Emparan, Alex Maloney, Victoria Martin, Shiraz Minwalla and Andrew Svesko for illuminating discussions. This work was supported in part by the US Department of Energy under grant DE-SC0019470.

\appendix

\section{Master equation}\label{A}
The linearized field equation is given by
\begin{equation}\label{6*}
\delta R_{\mu\nu}-\frac{R^0}{2}h_{\mu\nu}- g^0_{\mu\nu}\delta R=-\frac{m^2}{2}h_{\mu\nu}
\end{equation}\\
The general form of $g^0_{\mu\nu}$ is
\begin{equation}
ds^2=g_{ab}(y)dy^{a}dy^{b}+r^2d\sigma^2_{n+1} \footnote{Note: n here is different from n in  \cite{Kodama:2003jz,Kodama:2000fa}}
\end{equation}\\
where $d\sigma^2_{n+1}=\gamma_{ij}(x)dx^{i}dx^{j}$ with constant sectional curvature $K$. In our case, from the metric \eqref{metriceqn}, $a,b$ are either $t$ or $r$ coordinates and $i,j$ are coordinates on the $S^{n+1}$ sphere with $K=1$.
\subsection{$SO(n+2)$ Tensor modes}
Following \cite{Kodama:2003jz,Kodama:2000fa}, for tensor modes, the metric perturbation is given by \\
\begin{equation}
h_{ab}=0, \hspace{2mm} h_{ai}=0, \hspace{2mm} h_{ij}=2r^2H_T\mathbb{T}_{ij} 
\end{equation}\\
where $\mathbb{T}_{ij}$ are the harmonic tensors satisfying
\begin{equation}\\
(\hat{\Delta}+k^2)\mathbb{T}_{ij}=0
\end{equation}
where $k^2=l(l+n)-2,\hspace{2mm}l=2,\cdots$, and $\mathbb{T}_{ij}$ satisfy following properties 
\begin{equation}
\begin{split}
\mathbb{T}_{i}^{i}&=0\\
\hat{D}_j\mathbb{T}_{i}^j&=0
\end{split}
\end{equation}
Here $\hat{\Delta}$ and $\hat{D}$ correspond to the Laplacian operator and covariant derivative, respectively, on only the $n+1$ $i,j$ coordinates.\\ 

Using Appendix B of \cite{Kodama:2000fa}, the $ab$ and $ai$ components of \eqref{6*} are trivially satisfied, and the $ij$ component of \eqref{6*} simplifies to\\
\begin{equation}
2\delta R_{ij}=-r^2 \Box\left(\frac{1}{r^2}h_{ij}\right)-(n+1)\frac{D^ar}{r}D_ah_{ij}-\frac{1}{r^2}\hat{\Delta}h_{ij}+2\left[\frac{n+1}{r^2}+\frac{f}{r^2}-\frac{\Box r}{r}\right]h_{ij}
\end{equation}\\
where $D_a$ is the covariant derivative and $\Box$ is the d'Alembertian on $a,b$ coordinates.
Simplifying this further gives
\begin{equation}
\delta R_{ij}=- r^2 \Box H_T-(n+1)rD^arD_a H_T+(k^2+2)H_T.
\end{equation}
Since $R^0h_{ij}=0$ and $\delta R =0$, \eqref{6*} reduces to
\begin{equation}
\Box H_T-\frac{(n+1)}{r} D^arD_a H_T+\frac{(k^2+2)}{r^2}H_T-m^2H_T=0.
\end{equation}
Substituting $\Psi=r^{n/2}H_T$ gives
\begin{equation}\label{PsiTensor}
\Box\Psi-\frac{V_T}{f}\Psi=m^2\Psi,
\end{equation}
where $V_T$ is given by \eqref{2.7}.

Plugging $\Psi(t,r)=e^{-i\omega t}\psi(r)$ gives
\begin{equation}
\frac{d}{d r}\left(f \frac{d\psi}{dr}\right)-\frac{V_T }{f}\psi+\frac{\omega^2}{f}\psi=m^2\psi.
\end{equation}

\subsection{$SO(n+2)$ Vector modes}
For vector modes, the metric perturbation is given by
\begin{equation}
h_{ab}=0, \hspace{2mm} h_{ai}=r f_a \mathbb{V}_{i}, \hspace{2mm} h_{ij}=2r^2H_T\mathbb{V}_{ij},
\end{equation}
where in the massless case $f_a$ and $H_T$ are gauge dependent variables which can be combined in terms of a single gauge invariant variable $F_a$ as described in \cite{Kodama:2003jz,Kodama:2000fa}:
 \begin{equation}
 F_a=f_a+\frac{r}{k}D_a H_T.
 \end{equation}
 Here, $V_i$ are vector harmonics defined by
\begin{equation}
(\hat{\Delta}+k^2)\mathbb{V}_{i}=0,
\end{equation}
with the properties
\begin{align}
\hat{D}_i\mathbb{V}^{i}&=0,\\
\hspace{1mm}\mathbb{V}_{i}^i&=0, \\
\left[\hat{\Delta}+k^2-(n+2)\right]\mathbb{V}_{ij}&=0,\\
\hat{D}_j\mathbb{V}_{i}^j &=\frac{k^2-n}{2k}\mathbb{V}_{i}.
\end{align}
where $k^2=l(l+n)-1,\, l\in \mathbb{Z}, \,l>0$, and the index sums are taken only over the $n+1$ spherical coordinates $i,j$.
 
Writing in terms of $F_a$ and $H_T$ the $ij$ component of \eqref{6*} becomes:
\begin{equation}
\frac{k}{r^{n+1}}D_a(r^{n}F^a)=m^2 H_T.
\end{equation}
Because we have added a mass, $H_T$ is no longer pure gauge (no gauge freedom remains).  However, we are not interested in the behavior of the $H_T$ mode, since it is not present when we return to $m=0$.  Accordingly, we set
\begin{align} 
F^a&=\frac{1}{r^{n}}\epsilon^{ab}D_b\Omega+\frac{m^2}{kr^n}h^a \label{Fdef}\quad \Longrightarrow\\
\frac{k}{r^{n+1}}D_a(r^{n}F^a)&=0+\frac{m^2}{r^{n+1}}D_ah^a \quad \quad \quad\Longrightarrow\nn\\
H_T&=\frac{1}{r^{n+1}}D_ah^a.
\end{align}
Under this choice, the modes we are interested in lie in $\Omega$, whereas $h^a$ contains the `gauge' modes.  These modes do interact, but as we will see the coupling is a suppressed term in the $1/n$ expansion. Of course a proper gauge fixing procedure at $m=0$ should consider the effect of the ghosts which would arise when choosing the gauge $H_T=0$, but this procedure should be done separately.  We discuss this consideration further in the conclusion, and for now focus on the $\Omega$ modes.

The $ai$ component of \eqref{6*} is
\begin{equation}
\frac{1}{r^{n+2}}D^b\left[r^{n+3}\left\{D_b\left(\frac{F_a}{r}\right)-D_a\left(\frac{F_b}{r}\right)\right\}\right]-\frac{k^2-n}{r^2}F_a=m^2F_a-\frac{rm^2}{k}D_a H_T,
\end{equation}
which under the definition \eqref{Fdef} becomes
\begin{align}
\label{9.}
 D^b&\left[r^{n+3}\left\{D_b\left(\frac{\epsilon_{ac}D^c\Omega+\frac{m^2}{k}h_a}{r^{n+1}}\right)-D_a\left(\frac{\epsilon_{bc}D^c\Omega+\frac{m^2}{k}h_b}{r^{n+1}}\right)\right\}\right]=\nn
 \\
& (k^2-n)\epsilon_{ac}D^c\Omega+m^2r^2\epsilon_{ac}D^c\Omega+(k^2-n)\frac{m^2}{k}h_a+\frac{m^4}{k}h_ar^2-\frac{m^2r^{3+n}}{k}D_a\left(\frac{D_bh^b}{r^{n+1}}\right).
\end{align}
Using 
\begin{equation}
\epsilon^{ed}\epsilon_{ab}=\delta^d_a\delta^e_d-\delta^e_a\delta^d_b,
\end{equation}
we can rewrite the antisymmetric term inside the curly brackets on the left hand side of
\eqref{9.}. We find
\begin{align}
 D^b&\left[r^{n+3}\epsilon^{ed}\epsilon_{ab}
 D_e\left(\frac{\epsilon_{dc}D^c\Omega+\frac{m^2}{k}h_d}{r^{n+1}}\right)
 \right]=
 \\\nn
& (k^2-n)\epsilon_{ac}D^c\Omega+m^2r^2\epsilon_{ac}D^c\Omega+(k^2-n)\frac{m^2}{k}h_a+\frac{m^4}{k}h_ar^2-\frac{m^2r^{3+n}}{k}D_a\left(\frac{D_bh^b}{r^{n+1}}\right).
\end{align}
We now move all of the $h$ terms to the right hand side.  Since we can commute the $\epsilon$s through the covariant derivatives, and since $\epsilon^{ed}\epsilon_{dc}=\delta^e_c$, 
we find
\begin{align}
 \epsilon_{ab}D^b&\left[r^{n+3}
 D_c\left(\frac{D^c\Omega}{r^{n+1}}\right)
 -
(k^2-n)\Omega-m^2r^2\Omega\right]
+2m^2r\Omega\epsilon_{ab}D^br 
 = 
 \\\nn
& -D^b\left[r^{n+3}\epsilon^{cd}\epsilon_{ab}
 D_c\left(\frac{m^2}{k}{h_d}{r^{n+1}}\right)
 \right]
 +(k^2-n)\frac{m^2}{k}h_a+\frac{m^4}{k}h_ar^2-\frac{m^2r^{3+n}}{k}D_a\left(\frac{D_bh^b}{r^{n+1}}\right).
\end{align}
Contracting both sides with $\epsilon^{ea}$ we have
\begin{align}
D^e&\left[r^{n+3}
 D_c\left(\frac{D^c\Omega}{r^{n+1}}\right)
 -
(k^2-n)\Omega-m^2r^2\Omega\right]
+2m^2r\Omega D^e r 
 = 
 \\\nn
& -D^e\left[r^{n+3}\epsilon^{cd}
 D_c\left(\frac{m^2}{k}{h_d}{r^{n+1}}\right)
 \right]
 +
\epsilon^{ea} \left[\frac{(k^2-n)m^2+m^4r^2}{k}h_a-\frac{m^2r^{3+n}}{k}D_a\left(\frac{D_bh^b}{r^{n+1}}\right)\right],
\end{align}
which can be rewritten as
\begin{align}
D^e&\left[r^2 \Box \Omega-(n+1)r\left(D_c r \right)D^c\Omega
-
(k^2-n)\Omega-m^2r^2\Omega\right]
+2m^2r\Omega D^e r 
 = 
\\\nn
& -D^e\left[r^{n+3}\epsilon^{cd}
 D_c\left(\frac{m^2}{k}{h_d}{r^{n+1}}\right)
\right]
 +
\epsilon^{ea} \left[\frac{(k^2-n)m^2+m^4r^2}{k}h_a-\frac{m^2r^{3+n}}{k}D_a\left(\frac{D_bh^b}{r^{n+1}}\right)\right].
\end{align}
Setting $\Omega=r^{(n+1)/2}\Psi$ in the above equation as well as doing some index manipulation, we find
\begin{align}
D_a&\left[r^{(n+5)/2} \left(\Box \Psi-
\frac{V_V}{f}\Psi-m^2\Psi
\right)\right]
+2m^2r^{(n+3)/2} \delta_a^r\Psi  
 = 
\\\nn
& -D_a\left[r^{n+3}\epsilon^{cb}
 D_c\left(\frac{m^2}{k}{h_b}{r^{n+1}}\right)
\right]
 +
\epsilon_{ac} \left[\frac{(k^2-n)m^2+m^4r^2}{k}h^c-\frac{m^2r^{3+n}}{k}D^c\left(\frac{D_bh^b}{r^{n+1}}\right)\right],
\end{align}
where $V_V$ is given as in \eqref{2.8}:
\begin{equation}\label{2.8again}
V_V=\frac{n^2 f}{4 r^2}\left[\left(1+\frac{2l}{n}\right)^2-\frac{1}{n^2}-3\left(1+\frac{1}{n}\right)^2\left(\frac{1}{r}\right)^n\right].
\end{equation}
Note that we have already set $r_0=1$ here for calculational convenience.

The equation of motion that we would expect for $\Psi$ when adding a mass term is now visible, in the form
\begin{equation}\label{EOMguess}
\left(\Box \Psi-
\frac{V_V}{f}\Psi-m^2\Psi
\right),
\end{equation}
except that we would expect to set this equation equal to zero directly.  The solution relies on realizing that the function $h$ should be set to cancel the remaining nonzero pieces; that is, if we choose to parameterize the false modes that become pure gauge in the $m=0$ limit by insisting that $h$ solve
\begin{align}\label{hsoln}
2m^2r^{(n+3)/2} \delta_a^r\Psi  
= &
-D_a\left[r^{n+3}\epsilon^{cb}
D_c\left(\frac{m^2}{k}{h_b}{r^{n+1}}\right)
\right]
\\\nn &+
\epsilon_{ac} \left[\frac{(k^2-n)m^2+m^4r^2}{k}h^c-\frac{m^2r^{3+n}}{k}D^c\left(\frac{D_bh^b}{r^{n+1}}\right)\right]
\end{align}
by itself, then the effect of the extra mode will be removed correctly.  Accordingly, we make this choice.  Note that the term $2m^2r^{(n+3)/2} \delta_a^r\Psi$, for fixed order in $n$, is of order $1/n$ relative to the terms in the main $\Psi$ equation; for example, consider the $r$ component of the covariant derivative of the mass term:
\begin{equation}
D_a\left[-m^2r^{(n+5)/2}\Psi\right]=-\frac{n}{2}m^2 r^{(3+n)/2}\Psi+\mathcal{O}(n^0).
\end{equation}
This term is order $n$ for $\Psi\sim \calO(n^0)$, whereas the $2m^2r^{(n+3)/2} \delta_a^r\Psi$ term is only order $n^0$.  Thus it appears consistent to first solve the equation \eqref{EOMguess} set to zero at leading order, and then solve the leading order equation for $h^a$ according to \eqref{hsoln}.  Next we would solve \eqref{EOMguess} at next to leading order, and then solve the next to leading order equation for $h$, and so on.

In any case, with the choice \eqref{hsoln}, we find 
\begin{equation}
D_a\left[r^{(n+5)/2}\left(\Box \Psi-
\frac{V_V}{f}\Psi-m^2\Psi
\right)\right]=0.
\end{equation}
Since this equation must be true for both $a=t$ and $a=r$, and we will apply boundary conditions that rule out constant solutions, we finally obtain the expected equation of motion,
\begin{equation}\label{EOMexpected}
\frac{d}{d r}\left(f \frac{d}{dr}\psi\right)-\frac{V_V \psi}{f}+\frac{\omega^2}{f}\psi-m^2\psi=0,
\end{equation}
where we have also plugged in $\Psi(t,r)=e^{-i\omega t}\psi(r)$.


\section{Larger $k$ modes}\label{B}
In this Appendix we justify the reason for discarding all modes with $k\geq2$, as mentioned in Section \ref{DecoupledVector}. 

To first order, the vector wave equation \eqref{3.7} reduces to
\begin{equation}\label{49}
\begin{split}
-(\rho-1)\frac{d^2\psi^{(1)}}{d\rho^2}-\frac{d\psi^{(1)}}{d\rho}+\frac{(\rho-3+(4\mu_{0}^2-1)\rho)}{4\rho^2}\psi^{(1)}+
2(\rho-1)\ln \rho \frac{d^2\psi^{(0)}}{d\rho^2}&+\\
\left(2\ln \rho+\frac{(\rho-1)}{\rho}\right)\frac{d\psi^{(0)}}{d\rho}+
\left(\frac{l}{\rho}-\frac{3}{2\rho^2}-\frac{(\rho-3)\ln \rho}{2\rho^2} +
\frac{2\mu_{0}\mu_1}{\rho}\right)\psi^{(0)}
=0,
\end{split}
\end{equation}
where $\psi^{(0)}$ is given by \eqref{29c}.

Solving \eqref{49} for a generic value of $\mu_0$ is challenging so we solve it for a few values of $k$.

We can then compute the leading order quasinormal frequencies by finding solutions to \eqref{49} for each value of $k$, or $\mu_0$, that satisfy boundary conditions \eqref{NormBC} and \eqref{24}. These frequencies are as follows:\\
\begin{equation}\label{53}
\begin{split}
\omega_{(0)}&=-i4/9 (k=2),\\
\omega_{(0)}&= i33/25 (k=3),\\
\omega_{(0)}&= i832/425 (k=4).
\end{split}
\end{equation}\\
Since these frequencies don't satisfy the condition \eqref{3.19}, we discard these modes. We expect the same to happen for all other values of $k\geq2$ although we haven't provided a rigorous mathematical proof; such a proof may be possible using the method of induction. Discarding all such modes also justifies the choice $\Delta$ as mentioned in Section \ref{DecoupledVector}.

\section{Scalar modes}\label{C}
For scalar modes, we don't derive the master equation from Section \ref{DecoupledVector}, because adding the mass in the action is overly complicated. Instead we add the mass directly to the massless equation of motion.  We expect this will give the correct order order of poles in $\Delta$ plane.  The equation of motion becomes
\begin{equation}   
\bigg(\mathcal{L}+U_S+\frac{(\rho-1)(4\mu^2-1)}{4\rho}\bigg)\psi=0,
\end{equation}
where $U_S=\frac{V_S-\omega^2}{n^2}$ and $V_S$ is the scalar potential given by
\begin{equation}\label{scalarpotential}
V_S=\frac{f(r)Q(r)}{4r^2\left(2\mu+\frac{(n+2)(n+1)}{\rho}\right)^2},
\end{equation}\\
where $\mu=(l+n-1)(l-1)$ and $Q(r)$ is defined as\\
\begin{equation}
\begin{split}
Q(r)&=\frac{(n+2)^2(n+1)^4}{\rho^3}+(4\mu(2(n+3)^2-11(n+3)+18)\\&+(n+2)(n^2-1)(n-3))\frac{(n+2)(n+1)}{\rho^2}-((n-3)\mu+(n+2)(n^2-1))\frac{12(n+1)\mu}{\rho}\\&+16\mu^3+4(n+3)(n+1)\mu^2.
\end{split}
\end{equation}

For scalar modes, we need to consider two different regimes in the near-horizon region, first when $\rho=\mathcal{O}(1)$ and second when $\rho=\mathcal{O}(n)$ (which is still within the near zone $\rho<< e^{n}$). In order to properly capture the physics where  $\rho=\mathcal{O}(n)$ we introduce another variable \\
\begin{equation}
\bar{\rho}=\frac{\rho}{n}.
\end{equation}\\
We can expand and solve the equations while keeping $\bar{\rho}=\mathcal{O}(1)$. The solutions at small $\bar{\rho}$ can be matched to those at $\bar{\rho}=\mathcal{O}(1)$ in the new overlap zone $1<<\rho<<n$.\\
When $\bar{\rho}=\mathcal{O}(1)$ the leading order potential is
\begin{equation}
U_S^{(0)}=\frac{1}{4}\frac{4(l-1)^2\bar{\rho}^2-12(l-1)\bar{\rho}+1}{(2(l-1)\bar{\rho}+1)^2}.
\end{equation}\\
The leading order equation is
\begin{equation}
-\bar{\rho}^2 \frac{d^2\bar{\psi}^{(0)}}{d\bar{\rho}^2}-\bar{\rho}\frac{d\bar{\psi}^{(0)}}{d\bar{\rho}}+(U_S^0 +\frac{4\mu_{0}^2-1}{4}) \bar{\psi}^{(0)}=0.
\end{equation}\\
The solution to the above equation is
\begin{equation}\label{57}
\begin{split}
\bar{\psi}^{(0)}&=A_1\bar{\rho}^{-\mu_{0}}\frac{1+2\bar{\rho}-2l\bar{\rho}-2\mu_{0}+4\bar{\rho}\mu_{0}-4l\bar{\rho}\mu_{0}}{1-2\bar{\rho}+2l\bar{\rho}}+\\&A_2\bar{\rho}^{\mu_{0}}\frac{1+2\bar{\rho}-2l\bar{\rho}+2\mu_{0}-4\bar{\rho}\mu_{0}+4l\bar{\rho}\mu_{0}}{1-2\bar{\rho}+2l\bar{\rho}}.
\end{split}
\end{equation}\\
The asymptotic boundary condition \eqref{NormBC} sets $A_2=0$. This is true for all $\mu_{0}$ except for $\mu_{0}=\pm 1/2$ (since the two behaviors $\bar{\rho}^{\pm\mu_{0}}$ are not separable), so we will deal with this case separately. At $\rho=\mathcal{O}(1)$, the leading-order equation takes the following form
\begin{equation}
-(\rho-1)^2\frac{d^2\psi^{(0)}}{d\rho^2}-(\rho-1)\frac{d\psi^{(0)}}{d\rho}+\frac{(\rho-1)(4\mu_{0}^2\rho+1)}{4 \rho^2}\psi^{(0)}=0.
\end{equation}
The general solution to the above equation is
\begin{equation}
\psi^{(0)}_T=C_1\sqrt{\rho}\hspace{1mm}P_n(2\rho-1)+C_2\sqrt{\rho}\hspace{1mm}Q_n(2\rho-1).
\end{equation}

Thus, the leading order solution satisfying the horizon boundary condition \eqref{23} sets $C_2=0$ and $C_1=1$,\\
\begin{equation}\label{58}
\psi^{(0)}(\rho)=\sqrt{\rho}\hspace{1mm}P_n(2\rho-1),
\end{equation}
where $n=\mu_{0}-1/2$.\\
So to fix $A_1$ we expand \eqref{57} around small $\bar{\rho}$ and match with  \eqref{58} expanded around large $\rho$. At large $\rho$  \eqref{58} becomes\\
 \begin{equation}\label{59}
 \begin{split}
 \psi^{(0)}(\rho)&=\rho^{-\mu_{0}}\left(\frac{\Gamma(-2\mu_{0})}{\Gamma^2(\frac{1}{2}-\mu_{0})}+\frac{1}{\rho}\frac{(1+2\mu_{0})\Gamma(-2\mu_{0})}{4\Gamma^2(\frac{1}{2}-\mu_{0})}+\mathcal{O}(1/\rho^2)\right)+\\
 & \rho^{\mu_{0}}\left(\frac{\Gamma(2\mu_{0})}{\Gamma^2(\frac{1}{2}+\mu_{0})}-\frac{1}{\rho}\frac{(2\mu_{0}-1)\Gamma(2\mu_{0})}{4\Gamma^2(\frac{1}{2}+\mu_{0})}+\mathcal{O}(1/\rho^2)\right).
 \end{split}
 \end{equation}\\
Matching \eqref{59} and \eqref{57} requires 
\begin{equation}
\mu_{0}=-k-\frac{1}{2},\hspace{2mm} k\in\mathbb{Z}_{\geq 1}\, .
\end{equation}\\
For all $k\geq1$, \eqref{58} takes the form \\
\begin{equation}\label{61}
\psi^{(0)}(\rho)=\sqrt \rho(a_k \rho^k+ a_{k-1} \rho^{k-1}+...+ a_1 \rho+ a_0 ),
\end{equation}\\
and expanding \eqref{57} around $\bar{\rho}=0$ takes the form\\
\begin{equation}\label{62}
\bar{\psi}^{(0)}(\bar{\rho})=A_1 \sqrt {\bar \rho}(c_k \bar{\rho}^k+c_{k+1} \bar{\rho}^{k+1}+ \rm{higher\hspace{1mm}order\hspace{1mm}terms}).
\end{equation}\\
It can be seen that \eqref{61} and \eqref{62} are impossible to match $\forall k\geq1$. 
Now let's deal with $\mu_{0}=\pm 1/2$ cases. For both cases the solution is of the form \\
\begin{equation}\label{63}
\bar{\psi}^{(0)}(\bar{\rho})=A_1\frac{\sqrt{\bar{\rho}}}{1-2\bar{\rho}+2l\bar{\rho}},
\end{equation}\\
and they match \eqref{58} at small $\bar{\rho}$.
However, as $\bar{\rho}\rightarrow\infty$, $\bar{\psi}^{(0)}(\bar{\rho})\rightarrow 1/ \sqrt{\bar \rho}$, so the only mode that should be allowed is $\mu_{0}=1/2$.\\
If we include the next-to leading order, the solution for $\psi_S(\rho)$ with the horizon boundary condition is\\
\begin{equation}\label{64}
\psi_S(\rho)=\sqrt \rho(1+\frac{1}{n}((1-2l-2\mu_1+2i\omega_{(0)})\hspace{1mm} \ln \sqrt{\rho}-2(l-1)(\rho-1)-i\omega_{(0)}\hspace{1mm} \ln(\rho-1)).
\end{equation}\\
The expansion of $\psi_S(\rho)$ at large $\rho$ gives\\
\begin{equation}\label{65}
\begin{split}
\psi_S(\rho)&=\sqrt \rho+\frac{1}{n}\big(\frac{i\omega_{(0)}}{\sqrt{\rho}}+2(l-1) \sqrt{\rho}-\sqrt{\rho}\hspace{1mm}(2l+2\mu_1-1)\ln \sqrt{\rho}-\\&(2l-1)\rho^{3/2}+\mathcal{O}(\rho^{-3/2})\big).
\end{split}
\end{equation}\\
The solution for $\bar{\psi}(\bar{\rho})$ with the condition that $\bar{\psi}(\bar{\rho})\rightarrow \rho^{-\mu_{0}}$ as $\rho\rightarrow\infty$\\
\begin{equation}\label{66}
\begin{split}
\bar{\psi}(\bar{\rho})&=A_1\Bigg(\frac{\sqrt{\bar \rho}}{1-2\bar{\rho}+2l\bar{\rho}}-\frac{1}{n}
\Big(\frac{5(l-1)\bar{\rho}-\mu_1+2(l-1)^2\bar{\rho}^2(2l+2\mu_1-3)}{2(l-1)\sqrt{\bar{\rho}}(1+2(l-1)\bar{\rho})^2}+\\&\frac{\sqrt {\bar \rho}(2l+2\mu_1-1)\ln \sqrt{\bar \rho}}{1+2(l-1)\bar{\rho}}\Big)\Bigg).
\end{split}
\end{equation}\\
Matching the leading order at small $\bar{\rho}$ with \eqref{65} requires $A_1=\sqrt{n}+A_2/\sqrt{n}$, and the expansion in 1/n becomes\\
\begin{equation}
\begin{split}
\bar{\psi}^{(0)}_S&(\bar{\rho})+\frac{1}{n}\bar{\psi^{(1)}_S}(\bar{\rho})=
\sqrt{\rho}+\frac{\mu_1}{2(l-1)\sqrt{\rho}}+\frac{1}{2\sqrt{\rho}(l-1)n}\Big(A2 \mu_1+\rho(l-1)\\&(2A_2-5-4\mu_1+ (2l+2\mu_1-1)\ln n)-\rho(l-1)(2l+2\mu_1-1)\ln \rho\\&-4(l-1)^2\rho^2\Big).
\end{split}
\end{equation}\\
Matching the leading order term with \eqref{65} requires $\mu_1=0$. This also tells that at sub-leading order there is no term $\propto 1/\sqrt{\bar{\rho}}$ as required by \eqref{65}.  We find that there is such a term hidden in $\bar{\psi}^{(2)}_S(\bar{\rho})$ which would also contribute to order $1/n$.\\
\begin{equation}
\frac{1}{n^2}\bar{\psi}^{(2)}_S(\rho/n)=\frac{1}{n}\frac{l^2-l+\mu_2-\omega_{(0)}^2}{2(l-1)\sqrt{\rho}}.
\end{equation}\\
Matching the $\propto 1/\sqrt{\bar{\rho}}$ term in \eqref{65} with the above expression gives\\
\begin{equation}
i \omega_{(0)}=\frac{l^2-l+\mu_2-\omega_{(0)}^2}{2(l-1)}.
\end{equation}\\
Thus $\omega_{(0)}=-(l-1)\bigg(i\pm\frac{\sqrt{l-1+\mu_2}}{l-1}\bigg)$.
Requiring \eqref{3.19} sets
\begin{equation}
\mu_2=l+2p-(l+p)^2.
\end{equation}
The poles occur at
\begin{equation}
\mu^*_S=\frac{1}{2}+\frac{1}{n^2}(l+2p-(l+p)^2)+\mathcal{O}(1/n^3),
\end{equation}
or when,
\begin{equation}
\Delta^*_S=\frac{1}{n}(l+2p-(l+p)^2)+\mathcal{O}(1/n^2).
\end{equation}
As we see, the poles corresponding to scalar modes occur at sub-leading order in the $\Delta$ plane.  We do not address the contribution of these poles to the overall graviton partition function, mostly because we do not have the correct expression along the lines of \cite{Yasuda:1983hk} for the total graviton partition function in terms of the gauge-independent formalism of \cite{Kodama:2003jz, Kodama:2000fa}.

\bibliography{QNM}


\end{document}